\documentclass[5p]{elsarticle}

\usepackage{subfig}
\usepackage{multirow}
\usepackage{booktabs}
\usepackage{amsmath}
\usepackage{flushend}
\usepackage{siunitx} %
\DeclareSIUnit{\belmilliwatt}{Bm}
\DeclareSIUnit{\dBm}{\deci\belmilliwatt}
\usepackage{xcolor}
\usepackage[flushleft]{threeparttable} %

\usepackage{hyperref}

\begin{document}
\title{Downlink Channel Access Performance of NR-U: Impact of Numerology 
and Mini-Slots
on Coexistence with Wi-Fi in the 5 GHz Band}

\author[AGH]{Katarzyna~Kosek-Szott}
\ead{kks@agh.edu.pl}
\author[UniPa]{Alice Lo Valvo}
\ead{alice.lovalvo@unipa.it}
\author[AGH]{Szymon~Szott\corref{cor1}}
\ead{szott@agh.edu.pl}
\author[UniPa]{Pierluigi Gallo}
\ead{pierluigi.gallo@unipa.it}
\author[UniPa]{Ilenia Tinnirello}
\ead{ilenia.tinnirello@unipa.it}
\address[AGH]{AGH University of Science and Technology, Krakow, Poland}
\address[UniPa]{University of Palermo, Palermo, Italy}
\cortext[cor1]{Corresponding author}


\begin{abstract}
Coexistence between cellular systems and Wi-Fi gained the attention of the research community when LTE License Assisted Access (LAA) entered the unlicensed band.
The recent introduction of NR-U as part of 5G introduces new coexistence 
opportunities because it implements 
scalable numerology (flexible subcarrier spacing and OFDM symbol lengths), and non-slot based scheduling (mini-slots), which considerably impact channel access. 
This paper analyses the impact of NR-U settings on its coexistence with Wi-Fi networks and compares it with LAA operation using simulations and experiments.
First, we propose a downlink channel access simulation model,
which addresses the problem of the dependency and non-uniformity of transmission attempts of different nodes, as a result of the synchronization mechanism introduced by NR-U. 
Second, we validate the accuracy of the proposed model using FPGA-based LAA, NR-U, and Wi-Fi prototypes with over-the-air transmissions. 
Additionally, we show that replacing LAA with NR-U would not only allow to overcome the problem of bandwidth wastage caused by reservation signals but also, in some cases, to preserve fairness in channel access for both scheduled and random-access systems. 
Finally, we conclude that fair coexistence of the aforementioned systems in unlicensed bands is not guaranteed in general, and novel mechanisms are necessary for improving the sharing of resources between scheduled and contention-based technologies. 

\end{abstract}

\noindent\parbox[t]{\textwidth}{
{\Large Please cite this paper as: 

Katarzyna Kosek-Szott, Alice Lo Valvo, Szymon Szott, Pierluigi Gallo, Ilenia Tinnirello, ``Downlink channel access performance of NR-U: Impact of numerology and mini-slots on coexistence with Wi-Fi in the 5 GHz band'', Computer Networks, Volume 195, 2021. \url{https://doi.org/10.1016/j.comnet.2021.108188}
}
}

\vspace{2cm}
\begin{verbatim}
@article{kosekszott2021downlink,
title = {Downlink channel access performance of NR-U: 
Impact of numerology and mini-slots on coexistence with Wi-Fi in the 5 GHz band},
journal = {Computer Networks},
volume = {195},
pages = {108188},
year = {2021},
issn = {1389--1286},
doi = {https://doi.org/10.1016/j.comnet.2021.108188},
url = {https://www.sciencedirect.com/science/article/pii/S1389128621002437},
author = {Katarzyna Kosek-Szott and Alice {Lo Valvo} and Szymon Szott 
and Pierluigi Gallo and Ilenia Tinnirello},
keywords = {Coexistence, LTE LAA, 5G NR-U, IEEE 802.11, Unlicensed bands},
}
\end{verbatim}
\clearpage

\maketitle

\section{Introduction}
IEEE 802.11 is the incumbent technology in license-free frequency bands, 
whereas LTE-based technologies, such as LTE License Assisted Access (LAA), are only recently being deployed in these bands. 
When accessing the channel, both Wi-Fi and LTE-based technologies aim to grasp as much radio resources as possible but they differ in their respective channel access methods.
Wi-Fi uses random (i.e., contention-based) access, in which nodes can transmit at any time, given that the channel is idle, and a random backoff procedure is used to avoid collisions.
LTE is based on scheduled access, also known as reservation-based access or contention-free access, in which node transmissions are scheduled by a central controller (base station) and no intra-network collisions occur for data transmissions.
Now, both access types are expected to coexist in unlicensed, shared bands.

\begin{figure}
\centering
\includegraphics[width=\columnwidth]{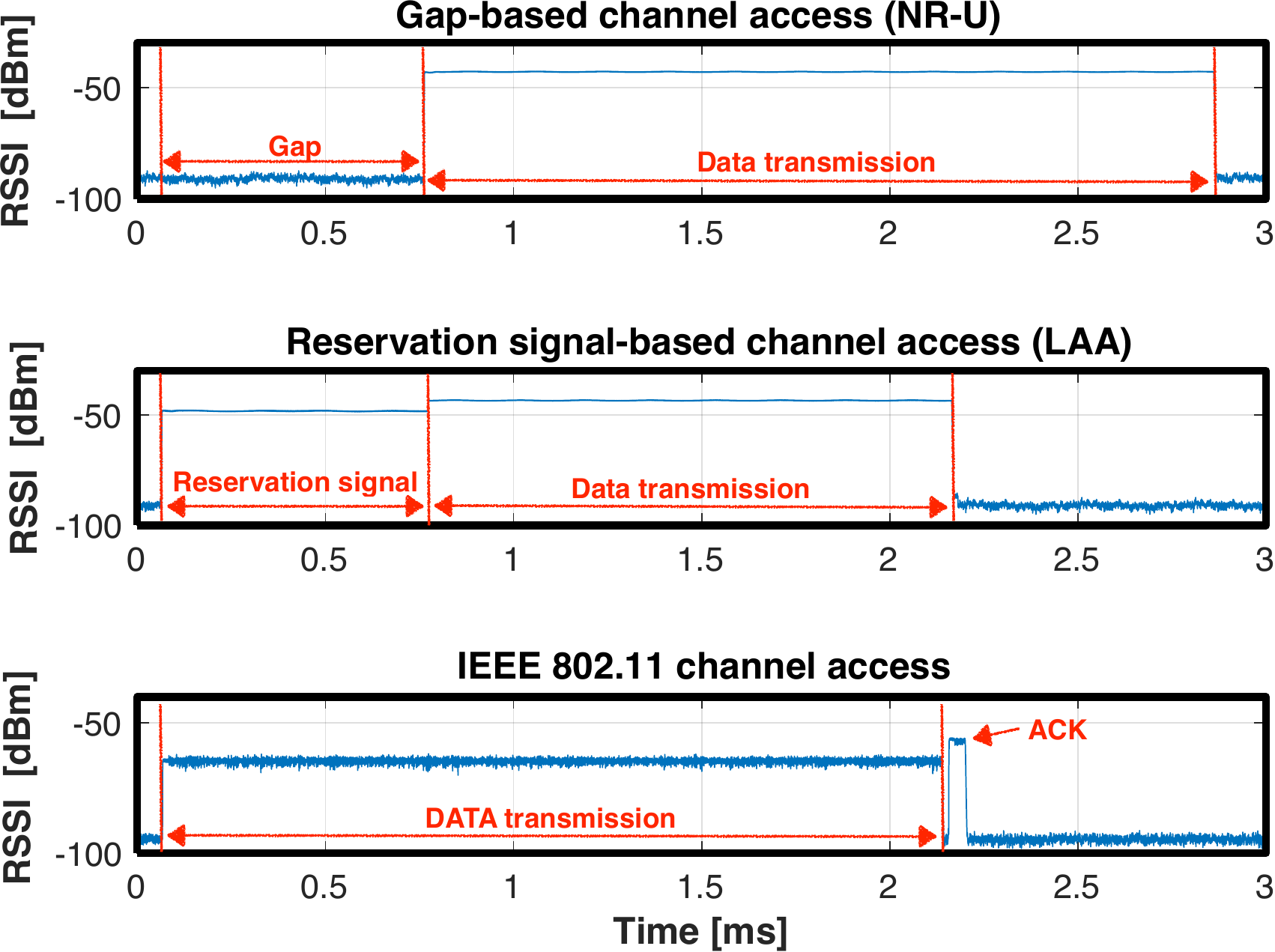}
\caption{Traces of over-the-air transmissions, captured by an USRP B210, from an exemplary testbed (described in Section~\ref{s:validation}) illustrating the implementation of various channel access mechanisms: gap (top), reservation signal (middle), and IEEE 802.11 (bottom). Note that the backoff countdown period is omitted for clarity of presentation.}
 \label{fig_sim}
\end{figure}

For use in unlicensed bands in Europe, LTE has had to conform to ETSI specification EN~301~893 \cite{etsi2017301}, which mandates a listen before talk (LBT) procedure to ensure all unlicensed radio access technologies have fair access to the channel in the 5~GHz band.
As a result, 3GPP has adopted LBT so that channel access for LAA is similar to that of Wi-Fi, through channel sensing and randomized backoff.
However, unlike Wi-Fi, LAA transmissions can only be scheduled to start at the beginning of an LTE slot boundary, which may not be synchronized with the end of the LBT procedure. %
This means that either (a) the channel access procedure has to be initialized after a self-deferral called the \textit{gap period} so that, if successful, it finishes at the beginning of a slot, or (b) channel access could begin immediately after the channel is idle but a \textit{reservation signal} (RS), also known as \textit{blocking energy}, would have to be transmitted to occupy the channel until the start of the next subframe. Thus, we have two alternative channel access mechanisms for LAA: gap-based and RS-based (Fig.~\ref{fig_sim}).
The RS approach has many disadvantages: it leads to a waste of radio resources\footnote{For 1~ms synchronization slots this means that only 1 bit is transmitted per 0.5 ms (on average) prior to each data transmission.} and has been criticized by both IEEE \cite{Nikolich2016} and researchers \cite{Loginov2018, Kutsevol2019, cierny2017fairness}.
ETSI's Broadband Radio Access Networks (BRAN) committee is currently working on an updated version of EN~301~893 which may disallow the use of RS \cite{Myles2019}.

While the operation of LAA remains undefined, 
the successor to LAA, New Radio--Unlicensed (NR-U) is being defined in 3GPP Release 16 as part of 5G. NR\nobreakdash-U partially solves the above-mentioned synchronization problem through its scalable numerology, i.e., flexible OFDM (orthogonal frequency-division multiplexing) symbol lengths, and scalable subcarrier spacing. This means that the granularity of channel access scheduling can be decreased from LAA's 1~\si{\milli\second} down to 125~\si{\micro\second}. Additionally, in NR-U, transmissions are not required to occupy 14 OFDM symbols (i.e., the whole subframe/slot duration), but can use mini-slots, improving scheduling granularity even down to 1 OFDM symbol \cite{zaidi20185g}.
We refer the reader to \cite{Patriciello2020, lagen2019new} for a more in-depth overview of the NR-U design. 

The impact of the scalable numerology introduced in NR-U on the coexistence with Wi-Fi has been analytically studied in \cite{Kutsevol2019Analytical} for a network scenario with a single LAA/NR-U base station and ideal sensing mechanisms. A simulation study based on ns-3, taking into account hardware limits and realistic antenna and channel models, has been proposed in \cite{Patriciello2020} for the mmWave (60~GHz) band, which has different challenges than the 5~GHz band.
For extending these results, after a description of the state of the art in Section~\ref{s:soa} and the unlicensed channel access rules in Section~\ref{sec:channel-access}, %
we provide the following novel contributions:
\begin{itemize}
    \item a unified simulation model for studying the coexistence of Wi-Fi and LAA/NR-U (Section~\ref{s:model}) contending on a single channel in the 5~GHz band, which addresses the problem of channel sensing delays and 
    non-uniformity of transmission attempts of different nodes over time, as a result of the synchronization requirement of NR-U;
    \item an experimental validation of the model using FPGA-based Wi-Fi, LAA, and NR-U prototypes supporting over-the-air transmissions (Section~\ref{s:validation}); 
    \item an upper bound performance analysis (Section~\ref{s:analysis}) based on collision probability and channel occupancy time (airtime) rather than throughput, to focus on fairness in channel access between different technologies and filter out the dependency on secondary aspects such as modulation or transmission rate, which can be different for the analyzed technologies and might unnecessarily complicate formulating conclusions.
\end{itemize}
We mostly focus on comparing Wi-Fi coexistence with LAA and NR-U implementing the RS-based and gap-based channel access mechanisms, respectively, and show the costs and benefits of both approaches. %
We additionally show that NR-U can avoid wasting bandwidth (by forgoing RSs) while, in some cases, preserving fairness in channel access for both scheduled and random-access systems. We also signalize that the introduction of additional mechanisms is necessary to provide fair coexistence of different network types in unlicensed bands. Finally, we focus on LAA/NR-U/Wi-Fi indoor dense coexistence scenarios, whilst outdoor scenarios are left for possible future study \cite{sathya2020measurement}.

\section{State of the Art in Coexistence Analysis}
\label{s:soa}
The coexistence of scheduled and random access wireless networks has been studied mostly for Wi-Fi, LTE LAA, LTE Unlicensed (LTE-U), and (recently) NR-U. 
LTE and Wi-Fi technologies are different in their contending logic, their sensing mechanism, and contention parameters. Therefore, methodologies applied to Wi-Fi performance analysis do not apply to LTE. 
In scheduled systems, transmission instants are correlated because of the synchronisation after the contention phase \cite{yi2017performance,Tinnirello2019}.
In conten\-tion-based systems, transmission instants are independent and are generally dealt with 2D or 3D Markov models, with the extra dimensions used for modeling additional features \cite{Xiao2019}.
As a result, the performance of scheduled and random access technologies are typically analysed separately from each other.
However, in the literature there are several papers which consider coexistence scenarios for Wi-Fi/LAA and Wi-Fi/NR-U.

\subsection{Wi-Fi/LAA Coexistence}
The coexistence of different channel access technologies is typically analyzed using models that start from either scheduled or contention-based channel access, and then relax some assumptions to include the other one.
For example, the authors of \cite{Pei2019,baswade2018modelling, Mehrnoush2018, hirzallah2019harmonious} use discrete time Markov chains to  model the performance of Wi-Fi and LAA coexistence. 
Unfortunately, slotted channel models based on the seminal work of Bianchi \cite{Bianchi2000} 
cannot be directly applied
for analyzing coexistence of random access with gap-based scheduled access technologies. 
Under such a synchronization mechanism, it is no longer true that each Wi-Fi transmission attempt may start at any channel time \cite{Tinnirello2019} (uniformity assumption) and that a given channel slot is idle independently of the state (idle or busy) of the previous channel slot.

To address temporal correlations between channel events over time, in \cite{Kutsevol2019Analytical} the authors model the whole interval between endings of two sequential successful Wi-Fi transmissions, rather than single idle or busy channel slots, under the assumption that collisions between Wi-Fi and LAA/NR-U nodes are not possible. 
Another approach to model coexistence is proposed in \cite{Cano2017}. The authors 
model the `cost of heterogeneity' (a result of either additional collisions introduced by LTE-U or the reservation signal introduced by LAA). They calculate the idle channel probability at periodic slot boundaries, considering random sampling by LAA nodes together with the NIMASTA property \cite{watabe2011analysis}, and show that this probability is relatively small, which serves as a justification of the introduction of the `cost of heterogeneity' by scheduled channel access systems. However, this model is simplified by having the duration of the reservation signal transmission set to a constant value (i.e., half of a synchronization slot length). Additionally, the `cost of heterogeneity' does not apply to gap-based access (NR-U). 

Furthermore, in \cite{gao2019lte} a solution to the fairness of LAA and Wi-Fi coexistence is proposed, based on adaptive contention window tuning and a careful selection of the maximum transmission duration duration. The authors focus on maximizing the total network throughput. However, such an approach is based on a constant synchronization delay for LAA %
and promotes non-standard behaviour.

In this paper we avoid modelling the coexistence of different radio access technologies based on Bianchi's work \cite{Bianchi2000}. Instead, we introduce additional parameters for modelling the system memory, in terms of absolute times and duration of the reservation signals and synchronization periods (Section \ref{s:model}). We focus mostly on the coexistence of Wi-Fi and NR-U, but we also show the most important differences in comparison to Wi-Fi/LAA coexistence. 

\subsection{Wi-Fi/NR-U Coexistence}
The problem of the coexistence of Wi-Fi and NR-U in the 5~GHz band is studied in \cite{Chen2019}. The authors propose a coexistence model focused on the PHY layer and the use of spatial multiplexing. They propose to coordinate the operation of Wi-Fi and NR-U by %
a 5G base station (gNB). To associate with the gNB, 802.11ax OFDMA and MU-MIMO transmissions are considered for Wi-Fi nodes and and 3GPP MU-MIMO transmission protocol is considered for NR-U nodes.

Coexistence issues arising between NR-U and Wi-Fi in the 6~GHz band have been described in \cite{Naik2020}, while \cite{Patriciello2020, verboom2020stochastic} provide coexistence studies for the mmWave (60~GHz) band. 
In particular, \cite{Naik2020} is a tutorial paper, in which challenges and opportunities for the next generation Wi-Fi and 5G NR-U in the 6 GHz bands are described.
Additionally, \cite{Patriciello2020} presents a simulation model developed for ns-3, which takes into account both LBT and duty cycle (LTE-U-like) channel access modes for NR-U. 
Furthermore, \cite{verboom2020stochastic} finds stochastic models for SINR and data rate under the assumption of downlink transmissions.

Finally, the problem of wireless network coexistence is being addressed by the industry, with recent patents on the physical layer for providing random access to NR-U \cite{pan2020physical} or methods for facilitating random access channels for 5G \cite{novlan2019facilitation}. 

We differentiate our work by focusing on the simulation and experimental validation of the coexistence of Wi-Fi and NR-U technologies in the 5 GHz band from the MAC-layer perspective. To the best of our knowledge, this area of research has not yet been addressed in the literature.

\section{Channel Access Rules in the 5~GHz Band}
\label{sec:channel-access}

Technologies operating in the 5~GHz unlicensed band in Europe must follow the rules provided in the ETSI EN 301 893 standard \cite{etsi2017301}, which defines the LBT mechanism for frame based equipment (FBE) and load based equipment (LBE). FBE uses a special transmit/receive structure with a fixed frame period whereas LBE uses load/demand driven transmissions/receptions. In this paper we focus only on LBE as it is used by both Wi-Fi and LAA/NR-U. Throughout the paper, whenever we refer to the LBT mechanism, we mean its LBE version.

\subsection{ETSI EN 301 893 Standard}
\label{sec:etsi}
Before transmitting, LBE waits for the channel to be idle for 16~\si{\micro\second} (which we refer to, by its Wi-Fi designation, as the short inter-frame space, SIFS). 
An idle channel is defined as having no other transmission detected above an energy detection threshold, set between $-75$ and $-85$~\si{\dBm/\mega\hertz}, depending on the equipment type. For equipment conforming to the 802.11-2016 standard \cite{802112016}, it is set to $-75$~\si{\dBm/\mega\hertz}, independently of the equipment's maximum transmit power. For other equipment types, it is proportional to the equipment's maximum transmit power.
While it is known that these energy detection threshold settings impact performance \cite{Mehrnoush2018}, in the remainder of this paper we assume all networks can perfectly detect each other, which allows us to focus only on contention in channel access.

Once the channel has been idle for 16~\si{\micro\second}, LBE performs clear channel assessment (CCA) in $p_0$ consecutive observation slots, each lasting at least 9~\si{\micro\second}. 
If the channel is occupied in any of these slots, the process restarts, whereas if the channel is idle for all $p_0$ slots, LBE starts the backoff procedure by
selecting a random number of observation slots $q \in \{0, 1, ..., CW\}$, where $CW$ is the contention window. CCA is performed for each observation slot: if the channel is idle, $q$ is decremented by one, otherwise the backoff contention is suspended and LBT is aborted. When $q=0$, LBE starts transmitting but this transmission cannot exceed the maximum channel occupancy time (MCOT) (Table~\ref{tab:params}). 
In summary, each channel access is preceded by a fixed ($p_0$) and random ($q$) number of slots. 
The latter depends on the current $CW$ value, which is reset to $CW_{min}$ after a successful transmission and doubled (up to $CW_{\max}$) before each retransmission.

Additionally, ETSI defines up to four priority classes. Each priority class has a different set of channel access parameter values (Table \ref{tab:params}). Furthermore, the $p_0$ and $CW_{\max}$ values may vary depending on transmission direction. 
To focus on understanding the basic principles of coexistence in channel access, we consider only downlink transmissions of best effort traffic: from the LAA/NR-U base station (eNB/gNB) to the user equipment (UE) or from the access point (AP) to Wi-Fi client stations. Uplink transmissions and other traffic categories are left for future study.

\begin{table*}[t]
  \centering
  \begin{threeparttable}
  \caption{Comparison of channel access parameter names and values for ETSI EN 301 893 \cite{etsi2017301}, 3GPP Rel-16 \cite{3gpp}, and IEEE 802.11 \cite{802112016} specifications. Priorities are listed from highest to lowers. AC -- Access category.}
  \label{tab:params}%
  \begin{tabular}{cccccccc}
    \toprule
    Specification & Priority & \multicolumn{1}{p{5em}}{Name of inter-frame space number parameter $p$} & \multicolumn{1}{p{5em}}{Value of inter-frame space number $p$ for DL (UL)} & \multicolumn{1}{p{3em}}{$CW_{\min}$} & \multicolumn{1}{p{4em}}{$CW_{\max}$ for DL (UL)} & \multicolumn{1}{p{8.5em}}{Name of maximum channel occupancy parameter $o_{\max}$} & \multicolumn{1}{p{7em}}{Value of maximum channel occupancy $o_{\max}$ for DL (UL) [\si{\milli\second}]} \\
    \midrule
    \multirow{4}[8]{*}{ETSI} & 4     & \multirow{4}[8]{*}{$p_0$} & 1 (2) & 3     & 7     & \multirow{4}[8]{*}{MCOT} & 2 \\
\cmidrule{2-2}\cmidrule{4-6}\cmidrule{8-8}          & 3     &       & 1 (2) & 7     & 15    &       & 4 \\
\cmidrule{2-2}\cmidrule{4-6}\cmidrule{8-8}          & 2     &       & 3     & 15    & 63 (1023) &       & 6\tnote{1, 2} \\
\cmidrule{2-2}\cmidrule{4-6}\cmidrule{8-8}          & 1     &       & 7     & 15    & 1023  &       & 6\tnote{1, 2} \\
    \midrule
    \multirow{4}[8]{*}{3GPP} & 1     & \multirow{4}[8]{*}{$m$} & 1 (2) & 3     & 7     & \multirow{4}[8]{*}{MCOT} & 2 \\
\cmidrule{2-2}\cmidrule{4-6}\cmidrule{8-8}          & 2     &       & 1 (2) & 7     & 15    &       & 3 (4) \\
\cmidrule{2-2}\cmidrule{4-6}\cmidrule{8-8}          & 3     &       & 3     & 15    & 63 (1023) &       & 8 (6)\tnote{1, 3} \\
\cmidrule{2-2}\cmidrule{4-6}\cmidrule{8-8}          & 4     &       & 7     & 15    & 1023  &       & 8 (6)\tnote{1, 3} \\
    \midrule
    \multirow{4}[8]{*}{802.11} & AC\_VO & \multirow{4}[8]{*}{AIFSN} & 1 (2) & 3     & 7     & \multirow{4}[8]{*}{$TXOP_{limit}$} & 2.08\tnote{4} \\
\cmidrule{2-2}\cmidrule{4-6}\cmidrule{8-8}          & AC\_VI &       & 1 (2) & 7     & 15    &       & 4.096\tnote{4} \\
\cmidrule{2-2}\cmidrule{4-6}\cmidrule{8-8}          & AC\_BE &       & 3     & 15    & 63 (1023) &       & 2.528\tnote{4} \\
\cmidrule{2-2}\cmidrule{4-6}\cmidrule{8-8}          & AC\_BK &       & 7     & 15    & 1023  &       & 2.528\tnote{4} \\
    \bottomrule
    \end{tabular}%
\begin{tablenotes}
  \footnotesize
  \item [1] For UL transmissions, MCOT can be extended up to 8 ms if one or more pauses with a minimum duration of 100~\si{\micro\second} are inserted.  
  \item [2] Additionally, MCOT can be increased up to 10 ms if $CW$ is doubled before and after each DL transmission exceeding 6 ms.
  \item [3] Additionally, MCOT can be extended up to 10 ms for both UL and DL transmissions in the absence of any other technology.
  \item [4] IEEE 802.11 also defines the maximum time of data transmission at the physical layer ($PPDUMaxTime$) equal (for all priorities) to 5.484~ms \cite{802112016}, which may be reached with frame aggregation \cite{kosek2019tuning}.
  \end{tablenotes} 
  \end{threeparttable}
\end{table*}%

\subsection{Comparison of IEEE 802.11 and 3GPP Channel Access}
The IEEE 802.11-2016 \cite{802112016} (Wi-Fi) channel access rules, which are part of its distributed coordination function (DCF), closely follow the specifications described in Section \ref{sec:etsi}.
Additionally, since 802.11 uses in-band acknowledgments for unicast frames, 
the recipient of a successfully transmitted frame may immediately (without LBT but within a SIFS) start transmitting an acknowledgement (ACK) frame. 

\begin{figure*}[t]
\centering
\includegraphics[width=\textwidth]{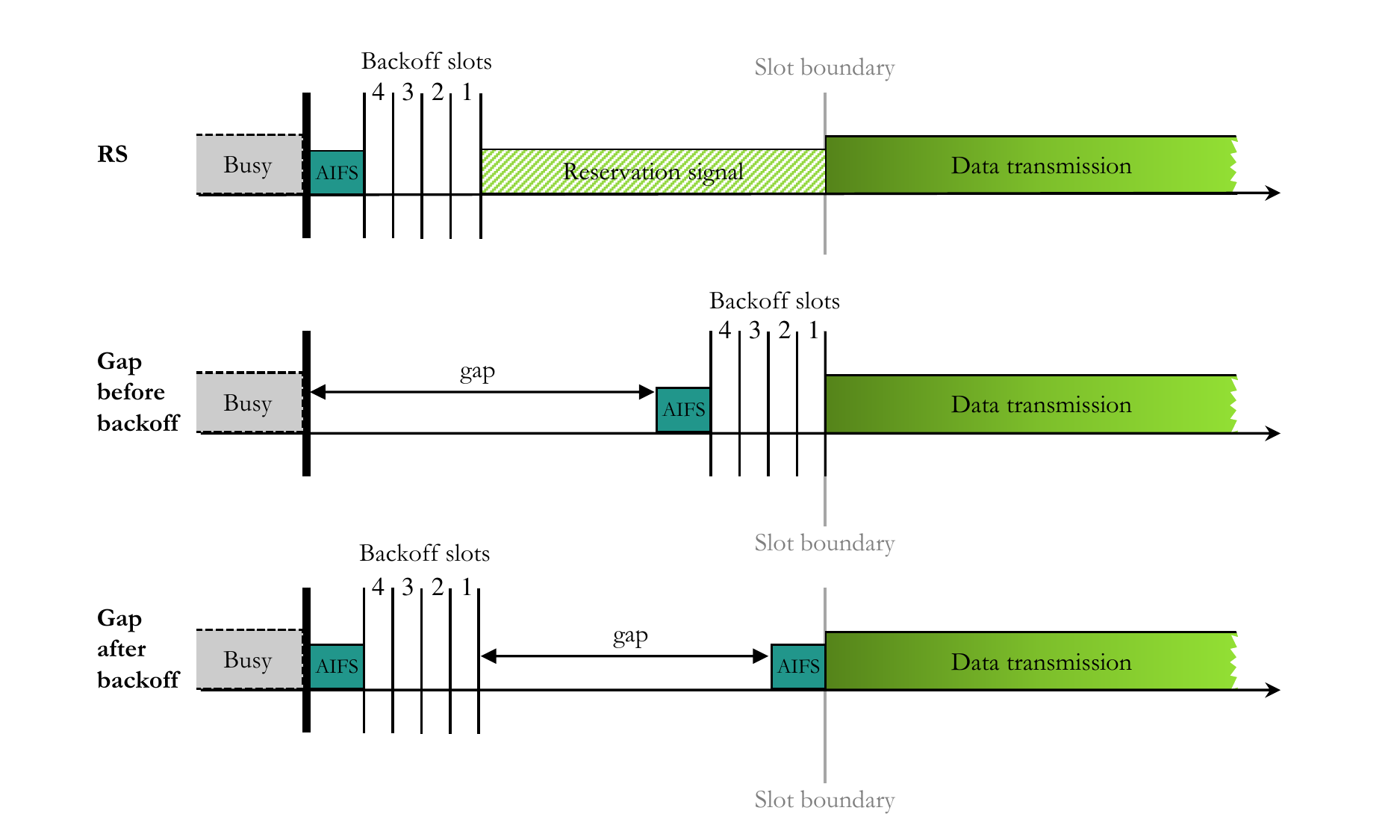}
\caption{A comparison of how LAA/NR-U can adapt their scheduled transmissions, which have to start at a slot boundary, to the random access nature of unlicensed channels: by transmitting a reservation signal (top) or by self-deferring with a gap period, either before (middle) or after (bottom) the backoff countdown. The AIFS blocks (named after 802.11's arbitration inter-frame spacing) are the fixed time intervals consisting of SIFS (16~\si{\micro\second}) and the fixed number of slots $p$ as required by LBT. In the \textit{gap after backoff} case, an additional check that the channel is idle is mandatory \cite{3gpp}.}
 \label{fig_gap_comparison}
\end{figure*}

From the MAC layer perspective, the operation of 3GPP LAA is similar to that of Wi-Fi, however, the acknowledgements are sent over the licensed band\footnote{LAA is based on carrier aggregation with devices having a connection in both the licensed and unlicensed bands \cite{Wszolek2020}.}. An important distinction from Wi-Fi is that LAA implements a mechanism to ensure that its transmission starts at a slot boundary \cite{Xiao2019}. For example, RSs are used after the backoff counter reaches zero to reserve the channel until the beginning of the slot boundary (Fig. \ref{fig_sim}) and the RS transmission duration counts towards the MCOT.
This operation was criticized by IEEE \cite{Nikolich2016} because it wastes channel resources and ``breaks underlying LBT sharing principles'' \cite{Myles2019}. 
Therefore, as an alternative to using an RS, an LAA node can perform a self-deferral by inserting a gap somewhere between the end of a previous transmission on the channel and the start of its own transmission.
As shown in Fig.~\ref{fig_gap_comparison}, there are at least two ways of implementing the gap mechanism (before or after backoff), both in line with \cite{3gpp}.
In the following, we assume the former approach as in \cite{mukherjee2016licensed}, while the latter approach has been used in \cite{cierny2017fairness,Xiao2019,Zheng2020}. 
Both approaches have their minor advantages and disadvantages with respect to performance, but we leave such analysis for future work.

The channel access procedures defined by 3GPP NR-U fully satisfy the rules defined in the ETSI specification (Section \ref{sec:etsi}). Therefore, the general NR-U operation seems similar to that of Wi-Fi with the 
difference that NR-U (like LAA) requires transmissions to start at a slot boundary. 
However, in contrast to LAA, NR-U introduces flexible slot duration and it is possible that it will implement only the gap mechanism (Fig.~\ref{fig_sim}), since the specification does not explicitly mention reservation signals \cite{3gpp}.
These changes may severely impact the observed performance, as we show in Section~\ref{s:analysis}, because the gap lengths result from the pre-configured slot lengths carefully selected by the operator (taking into account different trade-offs such as throughput vs. latency or slot length vs. switching overhead) and cannot be freely optimized to improve the NR-U network performance while coexisting with Wi-Fi.

In light of the above, in this paper we compare the \emph{coexistence of Wi-Fi with}: (i) \emph{LAA implementing the RS mechanism} and (ii) \emph{NR-U implementing the gap mechanism}. To avoid unnecessary repetitions, we omit the \emph{LAA with gap mechanism} case because it is covered by the \emph{NR-U with gap mechanism} case for a synchronization slot of 1~ms. 

\subsection{Comparison of IEEE 802.11, 3GPP, and ETSI Channel Access Parameters}
IEEE 802.11 and 3GPP, similarly to ETSI, define up to four priority classes with a similar set of channel access parameters. However, their naming partially differs for different standards as shown in Table~\ref{tab:params}. In the rest of the paper, to avoid ambiguity, we use $p$ to denote inter-frame space number and $o_{\max}$ to denote the maximum channel occupancy time. 

As shown in Table~\ref{tab:params}, $o_{\max}$ values differ among  technologies, which impacts the observed coexistence behaviour (Section~\ref{s:validation}). Furthermore, the IEEE 802.11-2016 standard defines not only the maximum channel occupancy time in the form of $TXOP_{limit}$ but also in the form of the maximum time of transmission at the physical layer ($o_{\max} =5.484$ ms), which can be reached, e.g., with frame aggregation \cite{kosek2019tuning} introduced in more recent Wi-Fi amendments (like 802.11n/ac/ax). Therefore, we use this $o_{\max}$ value to compare Wi-Fi and LAA/NR-U upper-bound operation in Section \ref{s:analysis}.

\subsection{3GPP Transmission Categories}
3GPP defines four transmission category types for LAA and NR-U (Cat.~1 to Cat.~4). Cat.~1 is used for immediate transmissions, i.e., when the time between two successive transmissions is no more than 16~\si{\micro\second}. Cat.~2 is used for transmissions preceded by LBT, i.e., when the time between two successive transmissions is between 16~\si{\micro\second} and 25~\si{\micro\second}. In a special case, for the transmission of a discovery reference signal (DRS) up to 1~ms long, Cat.~2 LBT is used to initiate the transmission attempt. Cat.~3 and Cat.~4 use LBT with a fixed or exponential random backoff, respectively. %
In this paper, we limit the analysis to Cat.~4 since it is most commonly used for data traffic transmission and for channel occupancy initiation, whereas we do not consider signalling such as DRS transmissions.

\subsection{3GPP Frame Structure}
LAA/NR-U transmissions are organized into 10~ms radio frames. 
Each frame is composed of 10 sub-frames (slots) each 1~ms long and consisting of 14 OFDM symbols. This design allows for slot-based scheduling. NR-U additionally implements flexible slot lengths, which scale with the new types of subcarrier spacing and symbol lengths (Table \ref{tab:nrparams}). 
Furthermore, NR-U introduces the concept of mini-slots for non-slot-based scheduling.
In Rel-15 (the first NR release) DL mini-slot duration was restricted to 2, 5, and 7 OFDM symbols. In Rel-16 (the second phase of NR) the mini-slot duration can last even a single OFDM symbol and, as a result, transmissions can start almost immediately after LBT to minimize the gap between the end of LBT and the start of the actual data transmission \cite{zaidi20185g}. In this paper, we assume the same operation for NR-U because, even though Rel-16 is under development, it is expected that, similarly to licensed NR, NR-U will support both slot-based and mini-slot-based resource allocation schemes \cite{ahmadi20195g}.

\begin{table*}
\centering
\begin{threeparttable}
\caption{NR-U parameters for user traffic transmissions}
\label{tab:nrparams}%
\begin{tabular}{p{20em}p{20em}}
\toprule
Parameter          & Value \\ \midrule
Slots per subframe $n_s$ & 1, 2, 4, or 8 \\
Subcarrier spacing & $15 \times n_s$ kHz \\
OFDM symbol duration & $66.67/n_s$~\si{\micro\second} \\
Cyclic prefix (CP) duration & $4.69/n_s$~\si{\micro\second} \\
OFDM symbol duration including CP & $71.35/n_s$~\si{\micro\second} \\
Number of OFDM symbols per slot & 14\tnote{1} \\
Frame duration & 10 ms \\
Sub-frame duration & 1 ms \\
Slot duration & $1000/n_s$~\si{\micro\second} \\
Mini-slot duration (3GPP Rel-15) & 2, 4, or 7 OFDM symbols ($\approx$ 18, 36, or 63~\si{\micro\second}) \\
Mini-slot duration (3GPP Rel-16) & 1 OFDM symbol ($\approx$ 9~\si{\micro\second}) \\
Synchronization slot duration $\Delta$ & 9, 18, 36, 63, 125, 250, 500, or 1000~\si{\micro\second}\\
\bottomrule
\end{tabular}%
\begin{tablenotes}
  \footnotesize
  \item [1] For $n_s=4$ and extended cyclic prefix, the number of OFDM symbols per slot is 12.
\end{tablenotes}
\end{threeparttable}
\end{table*}%

\section{Simulation Model of Channel Access}
\label{s:model}
In this section, we describe a single channel access contention model, taking into account the channel access procedures of Wi-Fi and LAA/NR-U.
The latter can attempt a data transmission only at the start of a specific synchronization slot. 
Therefore, even when these synchronization slots have the smallest possible size (i.e., a single OFDM symbol\footnote{A single OFDM symbol lasts $\approx 9$~\si{\micro\second} for 120 kHz sub-carrier spacing, which is equal to the duration of a single Wi-Fi slot for the OFDM PHY.}) and the channel access parameters are uniform for both technologies, the LAA/NR-U channel access operation is not equivalent to the Wi-Fi one.  
\begin{figure*}
\centering
\includegraphics[width=\textwidth]{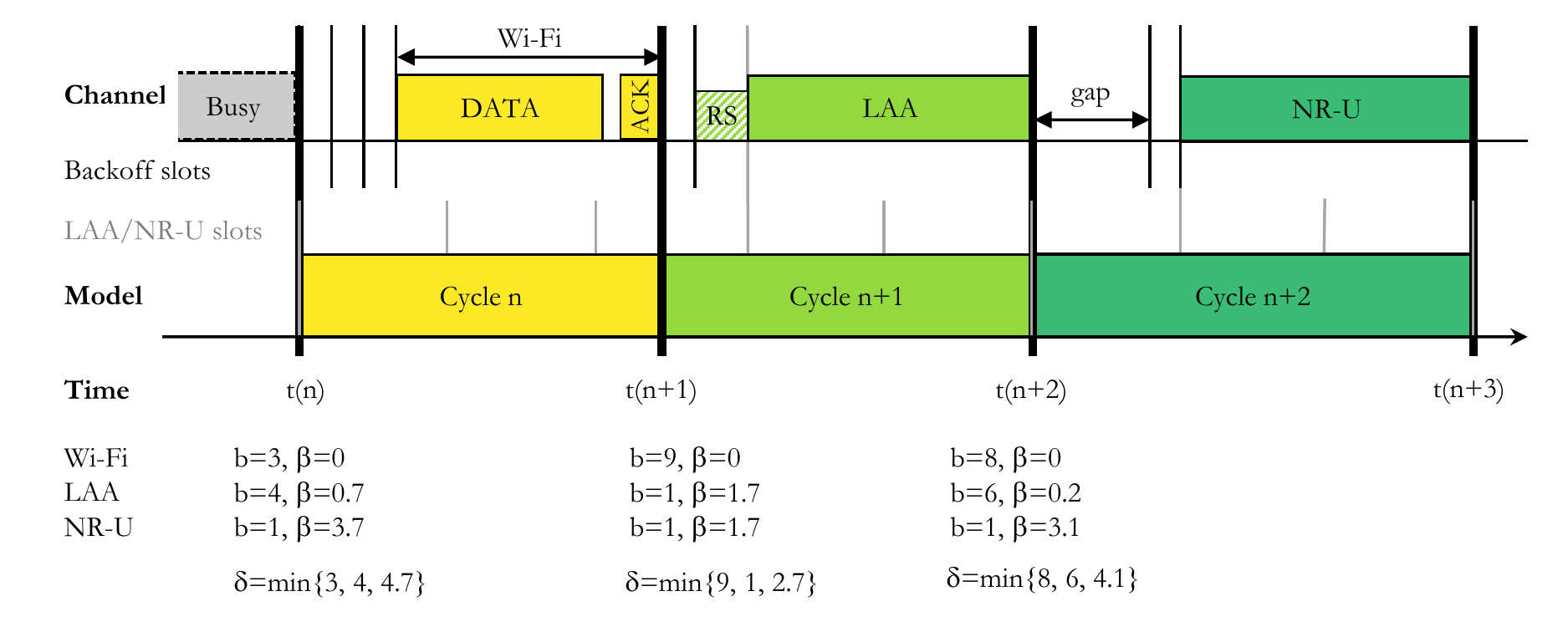}
\caption{An example contention between one Wi-Fi, one LAA, and one NR-U node. For clarity of presentation, we have removed the fixed time interval occurring at the start of each contention period $t$ and consisting of SIFS (16~\si{\micro\second}) and the fixed number of slots $p$ as required by LBT.}
 \label{fig_model}
\end{figure*}
We propose to model the different contention types of Wi-Fi and LAA/NR-U nodes by following the evolution of their backoff counters and additional synchronization times after each transmission attempt. 
By additional synchronization times we mean 
either the \textit{time spent for sending RSs after LBT} in case of LAA nodes or 
\textit{the gap time} spent as additional idle time by NR-U nodes. In both cases the synchronization times allow nodes to wait for the next available synchronization slot to begin, before the actual data transmission may take place. 
Fig.~\ref{fig_model} shows a channel access example in a network with one Wi-Fi, one LAA and one NR-U node, to clarify the impact of the additional synchronization time on the contention process. For readability, in the figure the synchronization slots are the same for both LAA and NR-U (same duration, fully synchronized\footnote{By fully synchronized we mean that the synchronization slots for all LAA/NR-U nodes are aligned. Throughout the remainder of this paper we assume that this is not the case, as it leads to a high collision rate among such nodes (which we demonstrate later on).}). LAA and NR-U nodes can start data transmission only at the occurrence of these synchronization slots, which in the figure are labeled as LAA/NR-U slots and drawn on the bottom of data transmissions.

\begin{table*}
  \centering
  \caption{Model nomenclature}
    \begin{tabular}{rl}
    \toprule
    Parameter & Description \\
    \midrule
    $\beta_k(n)$     & Additional time added to backoff countdown before node $k$ can transmit data  (i.e., gap or RS component)\\
    $b_k(n)$    & Backoff counter of a generic node $k$ at time $n$ \\
    $C$     & Collision probability \\
    $CW_{\max}$     & Maximum contention window \\
    $CW_{\min}$     & Minimum contention window \\
    $\delta$ &  Time required for starting a transmission, expressed in backoff slots\\ 
    $\Delta$     & Synchronization slot duration  \\
    $CS$     & Carrier sensing time \\
    $\gamma_k$     & Average synchronization time for node $k$ \\
    $\xi$    & Set of nodes winning the current contention \\
    $D$     & Data transmission duration \\
    $\lambda$     & Number of channel contentions \\
    $N$     & Total number of nodes \\
    $N_L$     & Number of LAA/NR-U nodes \\
    $N_W$     & Number of Wi-Fi nodes \\
    $O$     & Normalized channel occupancy time \\
    $O^T$     & Normalized total channel occupancy time \\
    $o_{\max}$     & Maximum channel occupancy time \\
    $p$     & Inter-frame space number \\
    $P_k$     & Channel occupancy time of node $k$ in case of transmission \\
    $q$    & Backoff counter \\
    $\sigma$ & Slot duration \\
    $S$     & Normalized successful channel occupancy time \\
    $S^\text{COT}$     & Normalized total channel occupancy time by technology \\
    $S^\text{EFF}$     & Normalized effective channel occupancy time \\
    $T$     &  Indicator of an additional synchronization time without transmitting (gap)\\
    \bottomrule
    \end{tabular}%
  \label{tab:nomenclature}%
\end{table*}%

\subsection{Formal Description of Access Mechanisms}
We assume nodes always have data to send (full buffer model) and we divide the channel access operations into consecutive cycles, encompassing a contention phase, during which the channel is idle so the nodes can decrease their backoff counters, and a channel occupancy phase, during which nodes attempt to transmit. Let $n$ be the discrete time corresponding to the end on the $n$-th transmission attempt. 
For modelling the contention mechanism, let $b_k(n)$ be the backoff counter of a generic node $k$ at time $n$; and let $\beta_k(n)$ be the additional synchronization time (measured in backoff slots): either RS for LAA or gap for NR-U. 
Let $N$ be the total number of contending nodes (among which there are $N_{\text{W}}$ Wi-Fi nodes and $N_{\text{L}}$ LAA/NR-U nodes, thus $N_{\text{W}}$+$N_{\text{L}}$=$N$).
We define the following additional vectors 
with $N$ components (Table~\ref{tab:nomenclature} lists the mathematical nomenclature used): 
\begin{itemize}
    \item $T$, whose generic $k$-th component is 1 when node $k$ spends an additional synchronization time $\beta_k$ without transmitting. This applies to NR-U nodes which employ the gap mechanism. For Wi-Fi and LAA, $T_k$ equals 0. 
    \item $P$, whose generic $k$-th component represents the channel occupancy time of node $k$ in case of transmission. 
    \item $p$, whose generic $k$-th component is given by the number of 
    slots required for resuming the backoff counter after 16~\si{\micro\second} of idle channel (SIFS) from the end of a previous transmission (i.e., the inter-frame space number). 
    \item $\Delta$, whose generic $k$-th component is node $k$'s synchronization slot: 0~ms for Wi-Fi nodes, 1~ms for LAA nodes (1 sub-frame), and a variable parameter for NR-U nodes ($\Delta \in \{9, 18, 36, 63, 125, 250, 500,\allowbreak 1000\}$~\si{\micro\second}, Table \ref{tab:nrparams}). 
\end{itemize}

The time required for starting a new transmission on the channel in the current contention,
expressed in the number of backoff slots, is equal to:

\begin{equation}
\delta(n)= \min\limits_{j=\{1,2, \cdots, N\}} \left\{ p_j+b_j(n) + T_j \cdot \beta_j(n) \right\},
\label{e:delta}
\end{equation}
where $\sigma$ is the slot duration of 9~\si{\micro\second} and $j$ is the node number.
Note that the contending nodes are not in general synchronized (i.e. $\delta(n)$ is not necessarily an integer number, given that $T_j \cdot \beta_j(n)$ are not an integer number for NR-U nodes employing the gap mechanism). 
When a new transmission is started on the channel, other nodes take some time to detect the energy, because of hardware limitations. Let $CS$ be the required sensing time, smaller than 0.5 backoff slots.
In case their backoff countdown is reset within this time, they also attempt to start a transmission.
Therefore, the set $\xi$ of nodes attempting a new transmission in the current contention is given by:
\begin{equation}
\xi = \{j: p_j + b_j(n) + T_j \cdot b_j(n) - \delta(n) < CS \}    
\end{equation}
which include the nodes starting a new transmission attempt (whose backoff countdown is completed exactly after $\delta(n)$), as well as nodes completing their backoff countdown before being able to detect the new transmission. 

If the set $\xi$ includes only one node, the contention results in a successful transmission, otherwise it results in a collision.
In Fig.~\ref{fig_model} there are three contention examples, in which $\delta(n)$ is calculated by comparing (i) the backoff counters of one Wi-Fi and one LAA node with (ii) the backoff counter plus the gap of one NR-U node. For the sake of figure readability, in the contentions we are not including $p$, which is assumed to be the same for all the nodes. 
During the gap, the channel is idle, therefore Wi-Fi and LAA nodes decrement their backoff counters by $\lceil \delta \rceil$ if they have data to transmit. 

It is easy to verify that the set of bi-dimensional parameters $\big(b_k(n), \beta_k(n)\big)$, for $k=\{1, 2, \dots, N\}$, represents the state of the system together with the scalar parameter $t(n)$ which maps the discrete time $n$ into an absolute time\footnote{Such an approach can be easily generalized to exponential backoff rules by simply adding a third state component $CW_k(n)$ representing the value of CW at time $n$ for node $k$. We do not introduce this component to simplify equations and focus on the analysis of the synchronization effects. However, we consider it in the Matlab implementation of our model (Section~\ref{s:matlab}).}

Both random and scheduled channel access mechanisms can be modeled by opportunistically defining the evolution of the system state at time $n+1$ as a function of the state at time $n$.
In particular, the evolution of the absolute time is given by
\begin{equation} t(n+1)=t(n)+\delta(n) \cdot \sigma + \max_{j \in C} P_j,\end{equation}
where $P_j$ is given by: 
$DATA+SIFS+ACK+SIFS$ if node $j$ is a Wi-Fi node and $DATA+SIFS$ 
if node $j$ is an LAA/NR-U node (as LAA/NR-U transmissions are acknowledged in the licensed band), with $DATA$ ($ACK$) being the time required for transmitting the data (acknowledgement) frame. 

As a result, for the backoff counter $b_k$ we get:
\begin{equation}
b_k(n+1) = \left \{ 
\begin{aligned}
&b_k(n) - \max\{ \lceil \delta(n) \rceil  -p_k , 0\}, k \notin \xi \\ 
&\qquad\text{for Wi-Fi/LAA,}\\
&b_k(n) - \max \{\lceil \delta(n)-\beta_k(n)\rceil - p_k, 0 \}, k \notin \xi \\ 
&\qquad\text{for NR-U,}\\ 
&rand(CW), k \in \xi,
\end{aligned}
\right.
\label{e:bk}
\end{equation}
where $\delta(n)$ is given by (\ref{e:delta}) and $rand(CW)$ is an integer extracted randomly in the range $[0, CW]$, with uniform distribution. 
Recall that, to clarify the presentation, we assume a fixed size contention window $CW$. For NR-U nodes we assume that after $p_k$ they wait for the additional synchronization time (gap) $\beta_k(n)$ before resuming their backoff counter
at the beginning of the $n+1$-th contention. It can be noticed that the length of this gap has a great impact on the backoff countdown, especially if the gap starts before the backoff countdown, as assumed in our work. The larger the synchronization slot (and, therefore, the probability of a large gap), the smaller is the probability of decreasing the backoff counter because of the higher probability of interfering transmissions. %

The additional synchronization time $\beta_k$ can be calculated as follows: 
\begin{equation}
\beta_k(n+1) = \left \{ 
\begin{aligned}
&0 \quad \text{for Wi-Fi,} \\
&\frac{\left \lceil \frac{\zeta}{\Delta_k}\right \rceil \times \Delta_k - \zeta}{\sigma} \quad \text{for LAA/NR-U,}
\end{aligned}
\right.
\label{e:RS}
\end{equation}
where $\zeta=t(n+1)+[p_k+b_k(n+1)]\cdot \sigma$ is the current time (in slots) combined with the fixed and random number of LBT slots. For LAA $\beta_k$ is the reservation signal component and for NR-U $\beta_k$ is the gap component. It can be noticed that their lengths are strongly dependent on the synchronization slot duration $\Delta_k$. 

Fig.~\ref{fig_model} summarizes the role of the contention parameters $b_k(n)$ and $\beta_k(n)$  in three consecutive contentions. The winning node is the one which may first start its transmission attempt: while Wi-Fi and LAA nodes resume backoff countdown after an idle time equal to $SIFS$ plus $p$ slots (omitted from the figure), NR-U needs to first wait for the gap interval $\beta$. 
Moreover, both LAA and NR-U nodes recompute their synchronization times at the start of each cycle, which vary when their residual backoff counters are different. 
In the first contention, the Wi-Fi node gains the channel and extracts a new backoff value at the end of its transmission. 
Meanwhile, the LAA node decreases its backoff counter by three slots but the NR-U node does not decrease its backoff at all
because of its additional gap. 
In the second contention, even though NR-U has the same backoff counter as LAA, the winner is still LAA, because its synchronization time $\beta=1.7$ is not a waiting time, but it is spent on transmitting the reservation signal. 
Since the LAA transmission is originated before the expiration of the NR-U gap, NR-U backoff is not decremented. 
In the last contention, the NR-U node can access the channel after an idle contention time equal to $b=1$ plus $\beta=3.1$ slots. 

\subsection{Performance Figures}
\label{s:perf_fig}
By monitoring the evolution of the system over time, we can derive the following relevant performance figures: 
\begin{itemize}
    \item normalized per-node channel occupancy time $O_k(\lambda)$ -- reflecting all the time node $k$ is transmitting (regardless of whether the transmission ended successfully or in a collision) normalized to the total time, 
    \item normalized per-node channel occupancy time for successful transmissions, normalized to the total time, in two variants: actual channel occupancy $S_k(\text{COT})$ and 
    effective time (without RS and ACKs) $S_k(\text{EFF})$, and  
    \item per-node collision probability ($C_k$). 
   \end{itemize}
We define these metrics in the following subsections.
\subsubsection{Normalized channel occupancy time}

For a given number of channel contentions $\lambda$, we are interested in monitoring the channel access time granted to each node. 
Let $x_k(n)$ be 1 if node $k$ is transmitting at time $n$, i.e., if $p_k + b_k(n) + T_k \cdot \beta_k(n)=\delta(n)$, and 0 otherwise. We define a normalized channel occupancy time $O_k(\lambda)$ for each node $k$ as
\begin{equation}
 O_k(\lambda) = \frac{ \sum_{n=0}^\lambda  x_k(n) \cdot  P_k }{\sum_{n=0}^\lambda t(n)}.
 \label{e:cot}
\end{equation}

Additionally, we calculate the total perceived channel occupancy time:
\begin{equation}
\begin{aligned}O^\text{T}(\lambda) =  O_W^\text{T}(\lambda) + O_L^\text{T}(\lambda) &= \sum_{k=1}^{N_\text{W}} O_k(\lambda) + \sum_{k=N_\text{W}+1}^{N} O_k(\lambda)\\
&=\sum_{k=1}^{N} O_k(\lambda),
\label{eq:tpcot}
\end{aligned}\end{equation}
where $ O_W^\text{T}(\lambda)$ is the total channel occupancy time for Wi-Fi nodes and $O_L^\text{T}(\lambda)$ is the total channel occupancy time for LAA/NR-U nodes.

\subsubsection{Normalized successful channel occupancy time}

We can define the normalized time $S_k(\lambda)$, considering only the channel access grants to node $k$ which results in a successful transmission. In such a case, we distinguish between the actual channel occupancy by technology (COT) and the effective time spent for transmitting data (EFF): 
\begin{equation}
S_k^{\text{COT}}(\lambda) = \frac{ \sum_{n=0}^\lambda s_k(n) \cdot P_k }{\sum_{n=0}^\lambda t(n)},
\end{equation}
\begin{equation}
S_k^{\text{EFF}}(\lambda) = \frac{ \sum_{n=0}^\lambda s_k(n) \cdot ( D_k - (1-T_k) \cdot \beta_k(n))}{\sum_{n=0}^\lambda t(n)},
\end{equation}
where $D_k$ is the data transmission duration equal to $DATA$ for Wi-Fi/LAA/NR-U nodes and $(1-T_k) \cdot \beta_k(n)$ is the time spent for transmitting the reservation signal for LAA nodes. 
Note that $D_k$ is at most equal to $o_{\max}$ (Table~\ref{tab:params}).
Additionally, we define $s_k(n)$ equal to 1 if node $k$ is the only one accessing the channel,  i.e., if  $p_k+b_k(n) +T_k \cdot \beta_k(n)=\delta(n)$
and $p_j+b_j(n) +T_j \cdot \beta_j(n)>\delta(n)$
for $j \neq k$, and 0 otherwise\footnote{This is a simplification because in case of a collision between transmissions of unequal duration, part of the data from the node transmitting the longest may still be decoded, e.g., if the LAA/NR-U transmission consists of several frames or Wi-Fi uses frame aggregation. We have omitted this aspect from the model to increase its clarity.}. 

Additionally, the total successful channel occupancy time $S^{\text{COT}}$ and the total effective successful channel occupancy time $S^{\text{EFF}}$ can be calculated as follows:
\begin{equation}
\begin{aligned}
S^{\text{COT}}(\lambda) &= S_{\text{W}}^{\text{COT}}(\lambda)+S_{\text{L}}^{\text{COT}}(\lambda)\\
&=\sum_{k=1}^{N_{\text{W}}} S_k^{\text{COT}}(\lambda) + \sum_{k=N_{\text{W}}+1}^{N}  S_k^{\text{COT}}(\lambda),
\end{aligned}
\end{equation}
\begin{equation}
\begin{aligned}
S^{\text{EFF}}(\lambda)&=S_{\text{W}}^{\text{EFF}}(\lambda)+S_{\text{L}}^{\text{EFF}}(\lambda) \\
&= \sum_{k=1}^{N_{\text{W}}}  S_k^{\text{EFF}}(\lambda) +\sum_{k=N_{\text{W}}+1}^{N}  S_k^{\text{EFF}}(\lambda), 
\end{aligned}
\end{equation}
where $S_{\text{W}}^{\text{COT}}(\lambda)$ ($S_{\text{L}}^{\text{COT}}(\lambda)$) and $S_{\text{W}}^{\text{EFF}}(\lambda)$ ($S_{\text{L}}^{\text{EFF}}(\lambda)$) are the Wi-Fi (LAA/NR-U) components.

Finally, note that $S_{\text{W}}^{\text{EFF}}(\lambda)$ and $S_{\text{L}}^{\text{EFF}}(\lambda)$ could be translated into Wi-Fi ($B_W$) and LAA/NR-U ($B_L$) effective throughput in the following way:
\begin{equation} B_W(\lambda) = r_W \cdot S_{\text{W}}^{\text{EFF}},
\label{eq:thr-wifi}\end{equation}
\begin{equation} B_L(\lambda) = r_L \cdot S_{\text{L}}^{\text{EFF}},
\label{eq:thr-nr}\end{equation}
where $r_W$ and $r_L$ are transmission rates for Wi-Fi and LAA/NR-U, respectively.

\subsubsection{Collision probability}

We define collision probability for node $k$ as
\begin{equation}C_k^\lambda=\frac{\sum_{n=0}^\lambda c_k(n)}{\sum_{n=0}^\lambda x_k(n)},\end{equation}
where $c_k(n)$ is 1 when node $k$ collides with another node at time $n$ and $c_k(n)=x_k(n)-s_k(n)$. 

\subsection{Model Implementation}
\label{s:matlab}
The model described above was implemented as a Monte Carlo simulator in Matlab based on the following observations.
First, since we are analyzing network devices each operating with a full-buffer traffic model, transmission attempts occur consecutively.
Second, since all devices are within hearability range (no hidden nodes), the end of a data transmission resets the channel state for all nodes (i.e., they all simultaneously begin to count the SIFS time, etc.).
Therefore, we can simulate channel access as a series of contention rounds, with each round equivalent to one cycle in Fig.~\ref{fig_model}.
Obviously contention rounds will vary in terms of duration (depending on the backoff slots preceding a transmission, the transmission duration, etc.) and this is taken into account by the simulator.

Within a single round, the simulator operates as follows.
First, it determines which nodes will win the contention by calculating the Wi-Fi/LAA nodes that have the lowest backoff ($b$) and the NR-U nodes that have the lowest backoff and gap ($b+\beta$), in accordance with (\ref{e:delta}).
Next, if the set of nodes which are determined to have won the contention consists of a single node, we consider this a successful transmission. Otherwise, there is a collision.
Furthermore, the simulator implements the channel access rules described in Section~\ref{sec:channel-access} including channel access parameters, backoff countdown, CW doubling, etc.
Once a sufficiently large number of contention rounds have elapsed, the simulation ends. 
Statistics are gathered throughout the course of the simulation to provide the performance figures mentioned in Section~\ref{s:perf_fig}.
Having defined the unified coexistence model, its performance metrics, and implementation, we can proceed with its experimental validation.

\section{Model Validation}
\label{s:validation}
To validate the accuracy of the simulation model we have implemented wireless node prototypes able to perform over-the-air transmissions and execute the LBT channel access procedure of Wi-Fi, LAA, and NR-U, including both gap and RS mechanisms.
Below, we describe the experimental setup, device configuration, scenario settings, and obtained results.

\subsection{Experimental Setup}
\label{sec:testbed}

Prototypes of custom wireless nodes are often built on top of software defined radio (SDR) platforms, such as the GNU Radio/USRP platform \cite{USRP}. 
We used the Wireless open-Access Research Platform (WARP) SDR platform \cite{WARP}, which includes a field-programmable gate array (FPGA)  circuit, to exploit both PHY layer programmability (to implement the LAA/NR-U and Wi-Fi PHYs) and MAC layer programmability (to implement the logic of LAA/NR-U and Wi-Fi MAC operation) inside the platform.
The 802.11 Reference Design developed by Mango Communication is already available in the used Xilinx Virtex-6 FPGAs:  OFDM transmissions are implemented in the PHY layer with multiple cores, whereas the MAC is implemented in software running on two MicroBlaze CPUs, with a support core in the FPGA to achieve accurate inter-frame timing. 
Three main blocks can be identified:
\begin{itemize}
    \item CPU -- of the two MicroBlaze CPUs the first executes the top-level MAC code and other high-level functions. All non-control packets for transmission and the various handshakes (probe request/response, association request/res\-ponse, etc.) are generated from this CPU, which is also responsible for implementing encapsulation and de-encapsulation of Ethernet frames according to the IEEE 802.11 standard. The second CPU executes the low-level code for the 802.11 MAC and is responsible for all MAC-PHY interactions and for handling intra-frame states including transmission of ACKs, scheduling of backoff counters, maintaining the contention window, and other DCF functions.
    \item MAC DCF -- this FPGA core acts as the interface between the MAC software and the TX/RX PHY cores and implements the timers required for DCF (timeout, backoff, SIFS, etc.) and the various carrier sensing mechanisms.
    \item PHY TX/RX -- these peripheral cores implement the OFDM physical layer transceiver specified in the 802.11 standard.
\end{itemize}

\begin{figure*}[t]
\centering
\includegraphics[width=\textwidth]{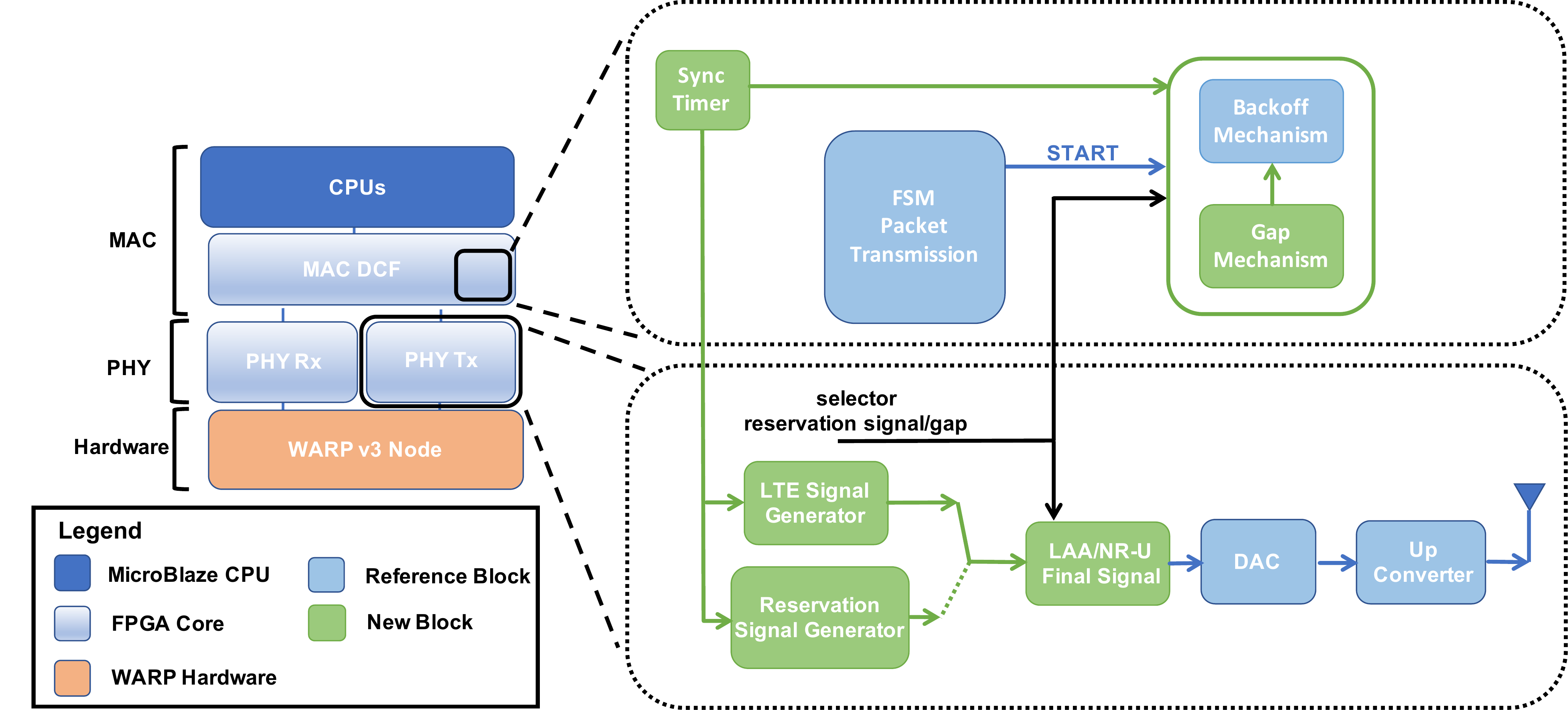}
\caption{Architecture and FPGA core modifications.}
 \label{fig_tx}
\end{figure*}

Fig.~\ref{fig_tx} shows the high-level architecture of the implementation of the 802.11 Reference Design with our modifications, which include the gap and reservation signal channel access mechanisms. The carrier sensing delay $CS$ of our implementation is 1~\si{\micro\second}, according to the datasheet of the  transceiver chip of our WARP board. For the generation of the final bitstream to program the FPGA we used Xilinx Platform Studio, while for changing the software in the MicroBlaze CPU and the PHY Blocks, we used the Xilinx Software Development Kit (SDK) and Xilinx System Generator, respectively.
In effect, for both versions we added functions in the CPUs to set register parameters responsible for the different node behaviour and to retrieve parameters from registers for the evaluation. The parameters are: synchronization slot ($\Delta$), data transmission length ($DATA$), and data transmission duration  ($D_k$). 
Moreover, we added a variable timer for introducing the synchronization slot $\Delta$ (the Sync Timer block) and an LAA/NR-U Signal Generator block in the MAC DCF and PHY TX, respectively. 
Specifically, since we are interested in a coexistence scenario, the LAA/NR-U Signal Generator block generates a sine signal able to change the state of the channel to busy. 
The ROM block consists of I/Q sine samples which are repeated multiple times until the required LAA/NR-U packet length is reached.

Finally, we added the following blocks for the correct functioning of each channel access method:
\begin{itemize}
\item \emph{Implementation of the gap mechanism.}
To maintain the synchronization of NR-U nodes, an additional gap time ($\beta$) was added before the backoff countdown in the MAC DCF Core. 
For the gap period, we used various Xilinx blocks such as Xilinx Adder, Xilinx Multiplier, and Xilinx Subtracter to perform the calculation of (\ref{e:RS}). It may be observed that a Xilinx Divider block is missing as division operations are problematic in FPGA. 
Unlike addition, subtraction, or multiplication, there is no simple logic operation that will generate a quotient because fixed-point operations do not produce a finite and predictable fixed-point result. 
To solve this problem, we used a Xilinx Multiplier with a factor of $1/\Delta$ as inferred by one of the CPUs.

\item \emph{Implementation of the reservation signal-based mechanism.}
In case of LAA nodes, we added a reservation signal in the transmission chain with the implementation of the Reservation Signal Generator block. 
This block comprises various counter and logical blocks to activate the generation, and read-only memory (ROM) with I/Q sine samples which generates a sinusoidal signal of duration equal to $\beta$, which is the remaining time until the beginning of the synchronization slot. 
The amplitude of the reservation signal is lower than the LAA data transmission to be easily visible during channel acquisition (as shown in Fig.~\ref{fig_sim}). After generation, the reservation signal is added before the LAA data transmission itself in the LAA Final Signal block. The entire signal is then ready to be transmitted by the transceiver.
\end{itemize}

\subsection{Device Configuration}
Regarding the configuration of the devices, each Wi-Fi transmitter is configured with the traffic generation rate and the $CW_{\min}$ and $CW_{\max}$ values. 
The traffic generator is an iperf client generating UDP packets of 1470 bytes (1500 bytes at the MAC layer) at such a rate that there are always packets to send (the full-buffer model). 
At the physical layer, the Wi-Fi nodes use the lowest possible transmission rate (6~Mb/s)\footnote{Note that the low transmission rate is not a problem as we do not use throughput as a metric but rather the channel occupancy time (airtime). This allows us to focus on fairness in channel access, filtering out the dependency on secondary aspects such as modulation or transmission rate.} to achieve the longest possible (for these devices) data transmission time of 2.1 ms. 
Conversely, on each LAA/NR-U transmitter it is possible to configure, apart from $CW_{\min}$ and $CW_{\max}$, the length of the data transmission (which we set either to 2.1 ms or 6 ms depending on the validation scenario), the channel access mode (i.e., gap or reservation signal), and the length of the synchronization slot $\Delta$ (set according to Table \ref{tab:nrparams}).

In terms of experimental output, for each Wi-Fi and LAA/NR-U node, we have the number of data transmissions per second and, for LAA nodes, also the average of the reservation signal length. 
This data is sufficient to proceed with the evaluation presented in Section~\ref{s:validation} because it allows to calculate $O_k$ in (\ref{e:cot}) for all technologies (Wi-Fi, LAA and NR-U) and $O^{\text{T}}$ to have the total perceived channel occupancy for all implementations.

\begin{figure*}[h]
\centering
\includegraphics[width=0.7\columnwidth]{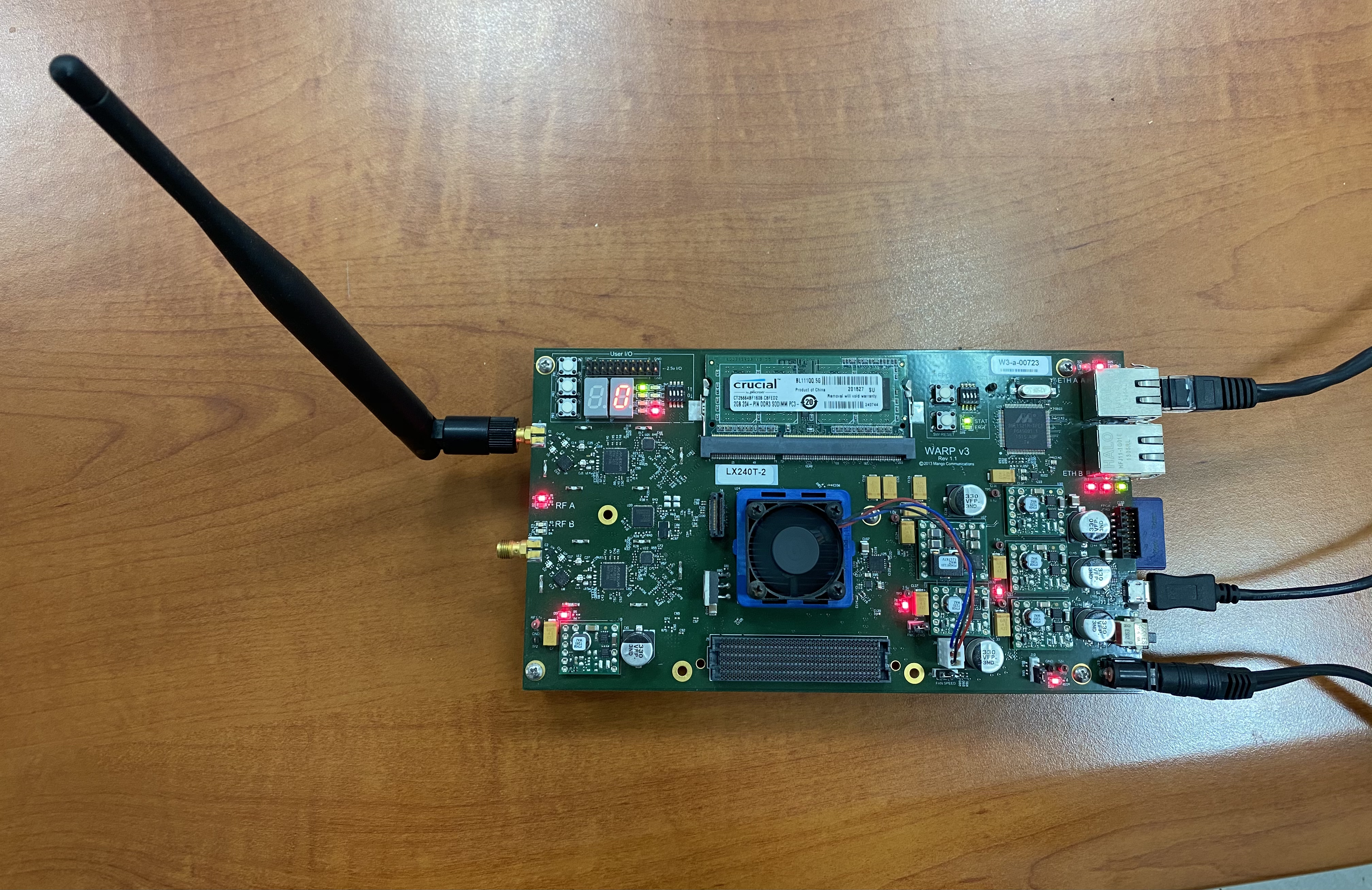}
\caption{An SDR WARP in the lab.}
 \label{fig_WARP}
\end{figure*}

We used five WARPs available in our lab, one is shown in Fig.~\ref{fig_WARP}, to set up different coexistence scenarios and we implemented up to two Wi-Fi links composed of a receiver and two transmitters, and up to two LAA/NR-U transmitters. All the transmissions of different coexistence scenarios were done over-the-air and inside a room of our lab. All the devices were placed in near proximity to each other in order to have direct links.

\subsection{Scenario Settings}
For validation we consider the following two coexistence scenarios:
\begin{itemize}
    \item \emph{Wi-Fi and LAA coexistence}. We analyze $O^\text{T}=O_W^\text{T}+O_L^\text{T}$,
    according to (\ref{eq:tpcot}),
    for a varying number of Wi-Fi and LAA nodes ($N_{\text{W}},N_{\text{L}}\in\{1,2\}$) sharing the wireless channel\footnote{In this section, for both simulations and experiments, we calculate $O^\text{T}$ (which includes all transmission attempts, i.e., both successful and collided transmissions). Therefore, the observed $O^\text{T}$ values can be greater than one. Recall that for LAA, the collisions are handled in the licensed bands and it is not always possible to distinguish collisions from successful transmissions while analysing the unlicensed channel. }. LAA uses RS-based channel access with a  synchronization slot set to 1000~\si{\micro\second}.
    \item \emph{NR-U and Wi-Fi coexistence}. We again analyze $O^\text{T}$ but for two NR-U nodes coexisting with two Wi-Fi nodes ($N_{\text{L}}=N_{\text{W}}=2$) under a varying duration of the synchronization slot for NR-U: $\Delta \in \{9, 18, 32, 63,\allowbreak 125, 250, 500, 1000 \}$~\si{\micro\second}. Only the gap approach is considered, i.e., no reservation signal is transmitted after a successful LBT procedure\footnote{Note that NR-U with $\Delta = 1000$~\si{\micro\second} corresponds to LAA with gap channel access.}.
\end{itemize}
In both scenarios either homogeneous (2.1~\si{\milli\second}) or heterogeneous (2.1~\si{\milli\second} for Wi-Fi and 6~\si{\milli\second} for LAA/NR-U) data transmission duration $D_k$ settings were used. 
The generated traffic was classified into the BE access category (802.11) or as priority 3 (LAA/NR-U), with the study of other priorities planned as future work. 
All other simulation parameters were set according to Tables~\ref{tab:params} and~\ref{tab:nrparams}.
Multiple runs were performed: 10 for the simulation model of Section~\ref{s:matlab} and three for the experiments.
The figures present the $95\%$ confidence intervals.

\begin{figure*}
\centering
\subfloat[Homogeneous frame length]{
\includegraphics[width=0.45\textwidth]{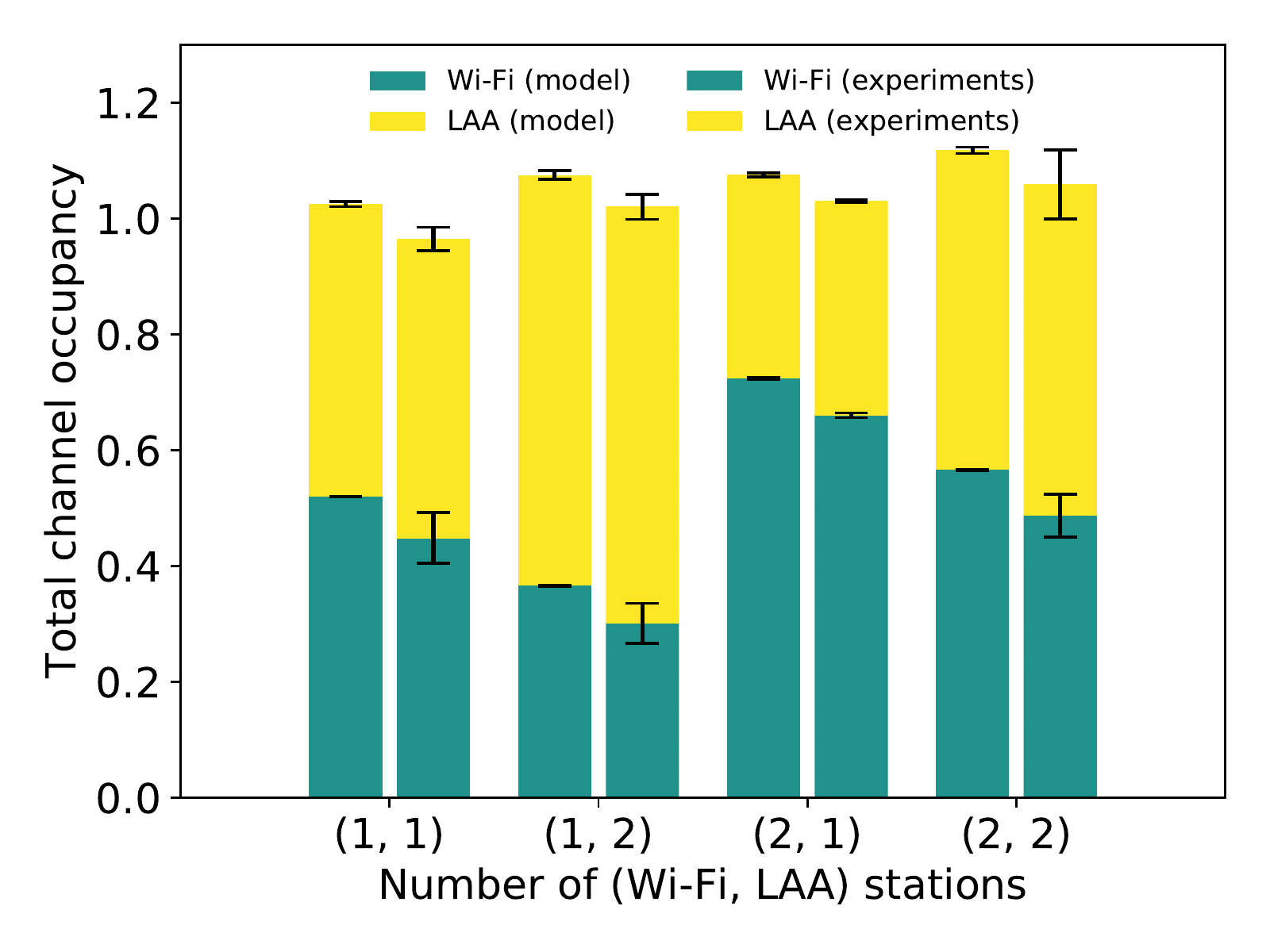}
\label{fig-validation-laa-2100}}
\hfil
\subfloat[Heterogeneous frame length]{
\includegraphics[width=0.45\textwidth]{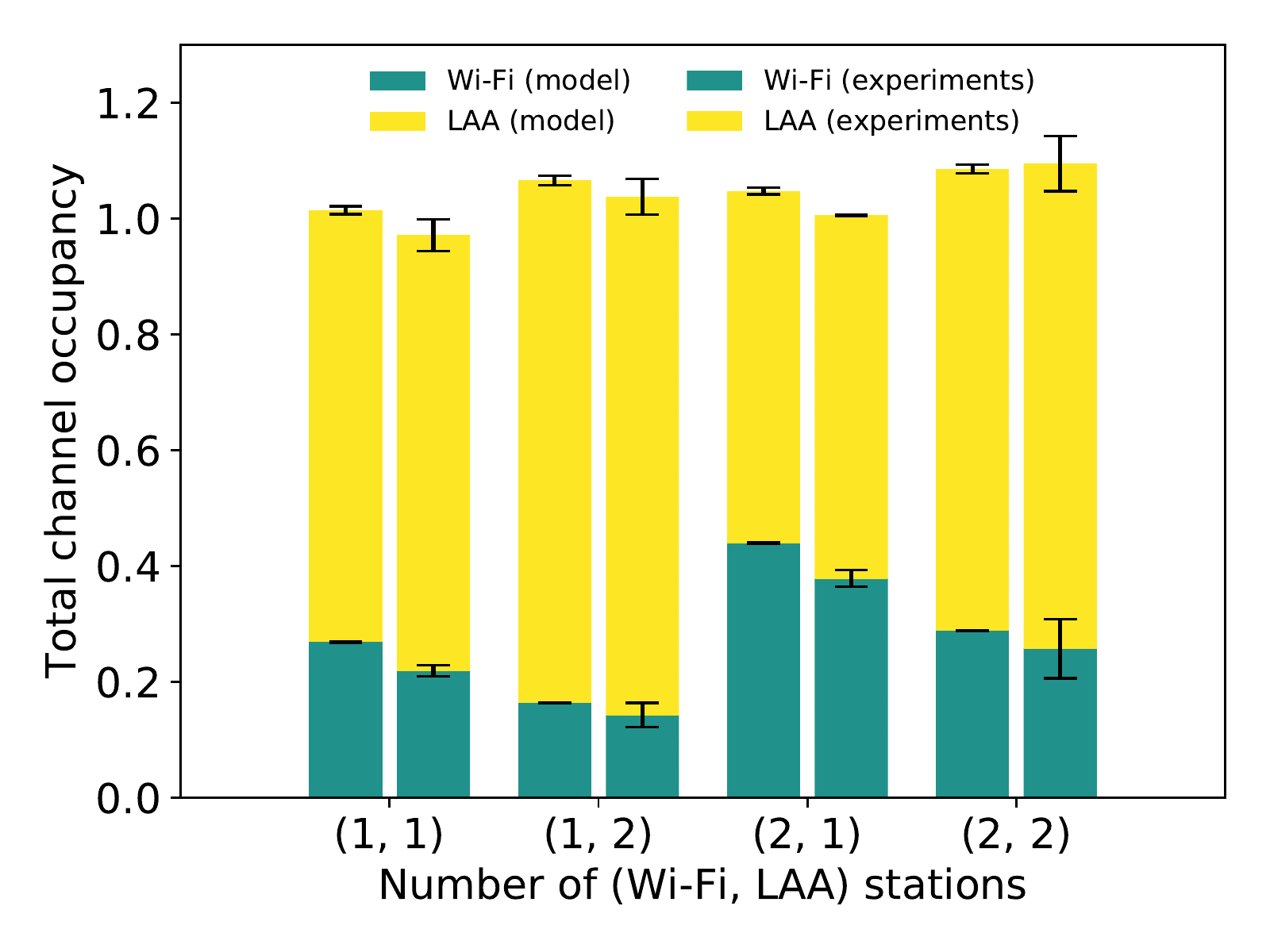}
\label{fig-validation-laa-6000}}
\caption{Validation results for Wi-Fi/LAA coexistence: comparison of normalized total channel occupancy time $O^T$ for various node configurations.}
\label{fig-validation-laa}
\end{figure*}

\subsection{Validation Results}
In the following we compare the simulation results of the model with the experimental results from the testbed.
We focus on comparing the results between both methods whereas a 
detailed discussion of the performance of each technology will be presented in Section~\ref{s:analysis}.

For the Wi-Fi/LAA scenario, Fig.~\ref{fig-validation-laa} shows a comparison between simulation (solid bars) and experimental (dash\-ed bars) results. 
For both homogeneous %
and heterogeneous %
frame lengths, $O^T$ levels match well between both simulation model and experiments. 
The error bars for experiments are larger due to the higher variability of channel conditions.

\begin{figure*}
\centering
\subfloat[Homogeneous frame length]{
\includegraphics[width=0.45\textwidth]{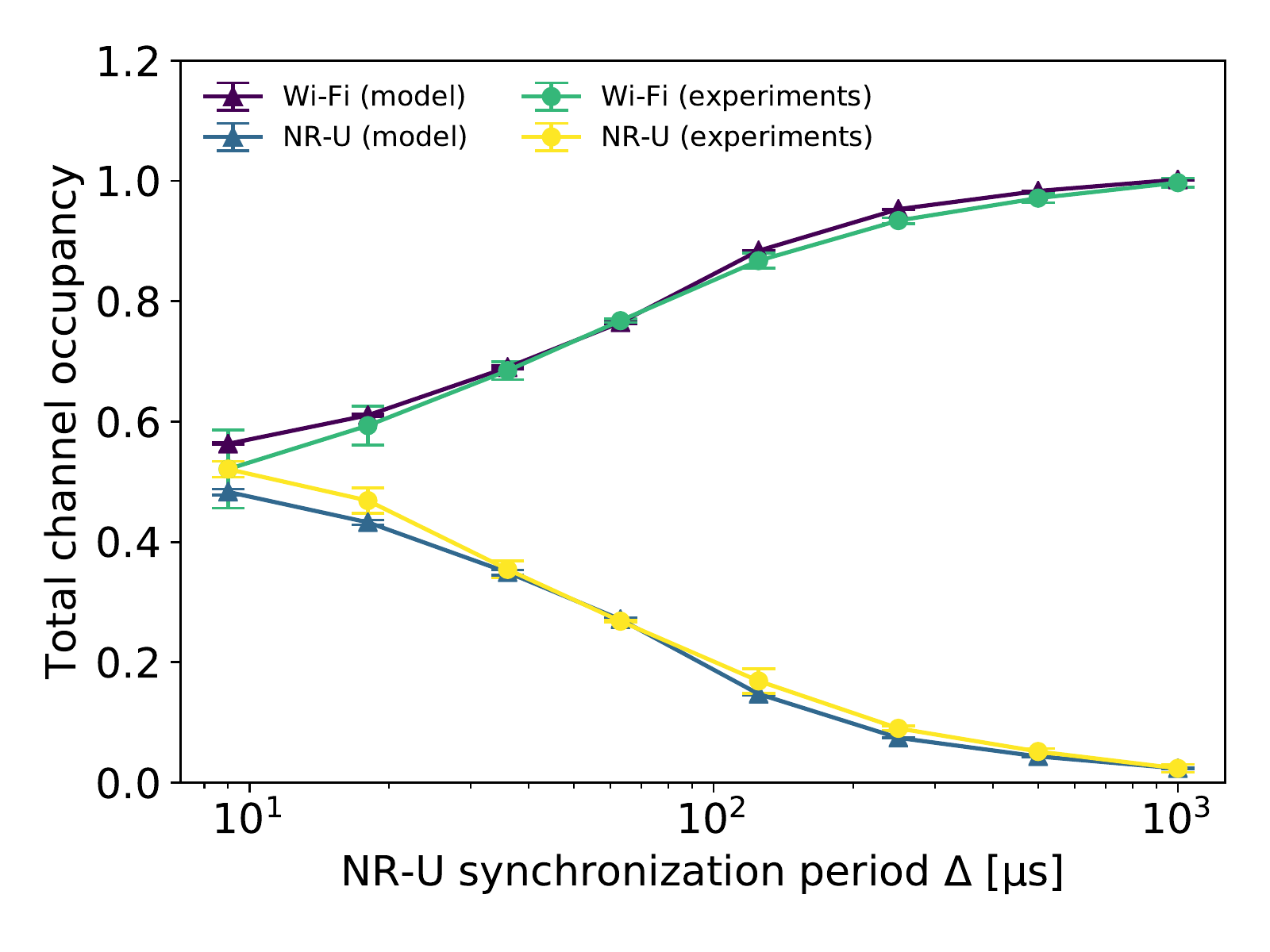}
\label{fig-validation-nr-2100}}
\subfloat[Heterogeneous frame length]{
\includegraphics[width=0.45\textwidth]{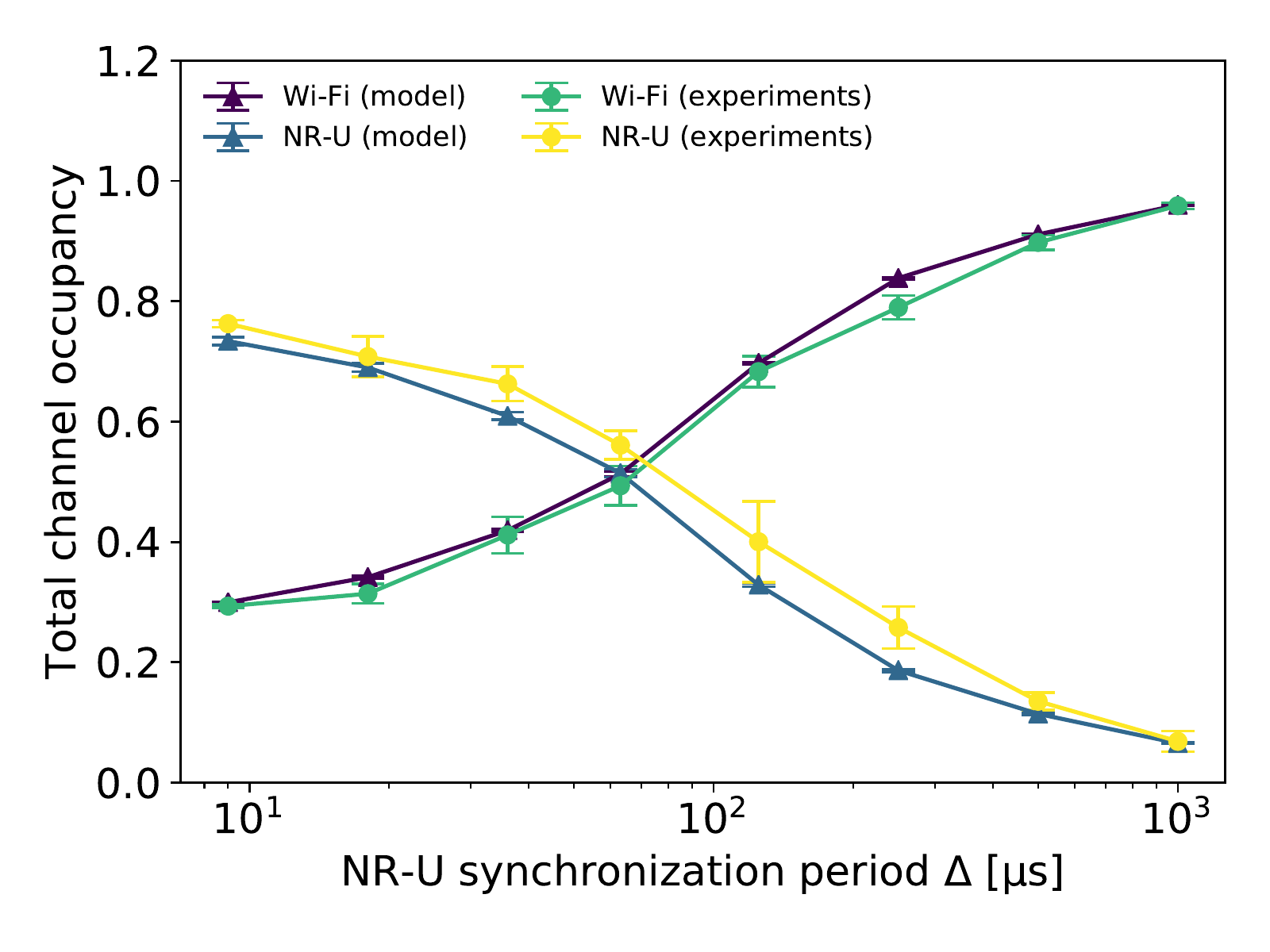}
\label{fig-validation-nr-6000}}
\caption{Validation results for Wi-Fi/NR-U coexistence: comparison of normalized total channel occupancy time $O^T$ for various synchronization slots.}
\label{fig-validation-nr}
\end{figure*}
A similar comparison is made for the Wi-Fi/NR-U scenario (Fig.~\ref{fig-validation-nr}). 
Simulation and experimental results resemble each other not only regardless of frame length but also for all considered NR-U synchronization slots.

The presented results validate our unified coexistence model in both Wi-Fi/LAA and Wi-Fi/NR-U scenarios. 
Therefore, in the next section we use this model to perform a simulation model-based analysis of more complex scenarios.

\section{Performance Analysis}
\label{s:analysis}
In this section we analyse the coexistence of Wi-Fi with LAA and NR-U nodes within an overlapping communication range (i.e., with no hidden nodes). In particular, we compare how the gap and the reservation signal-based channel access mechanisms impact Wi-Fi operation in the 5~GHz unlicensed band. Additionally, we show how different synchronization slot lengths impact channel shares for Wi-Fi and LAA/NR-U. We also answer the question who is a better neighbor when Wi-Fi and LAA/NR-U compete in the same channel. In all scenarios we assumed the default $CW$ and $p$ settings for the BE traffic, the maximum channel occupancy time allowed (i.e., the upper-bound) for each technology (Table~\ref{tab:params}): 6~ms for LAA/NR\nobreakdash-U and 5.4~ms for Wi-Fi, omni-directional antennas, and fixed-width channels with an all-or-nothing reservation approach. 
By calculating channel occupancy time (rather than throughput) our analysis is channel width agnostic. Furthermore, to establish upper performance bounds, we consider only data and (in case of Wi-Fi, cf. Section \ref{s:model}) acknowledgement transmissions, neglecting all other 802.11 control frames and LAA/NR-U's discovery reference signals. 

\subsection{Wi-Fi/LAA Coexistence}

\begin{figure*}
\centering
\subfloat[Channel occupancy time for RS-based access]{
\includegraphics[width=0.45\textwidth]{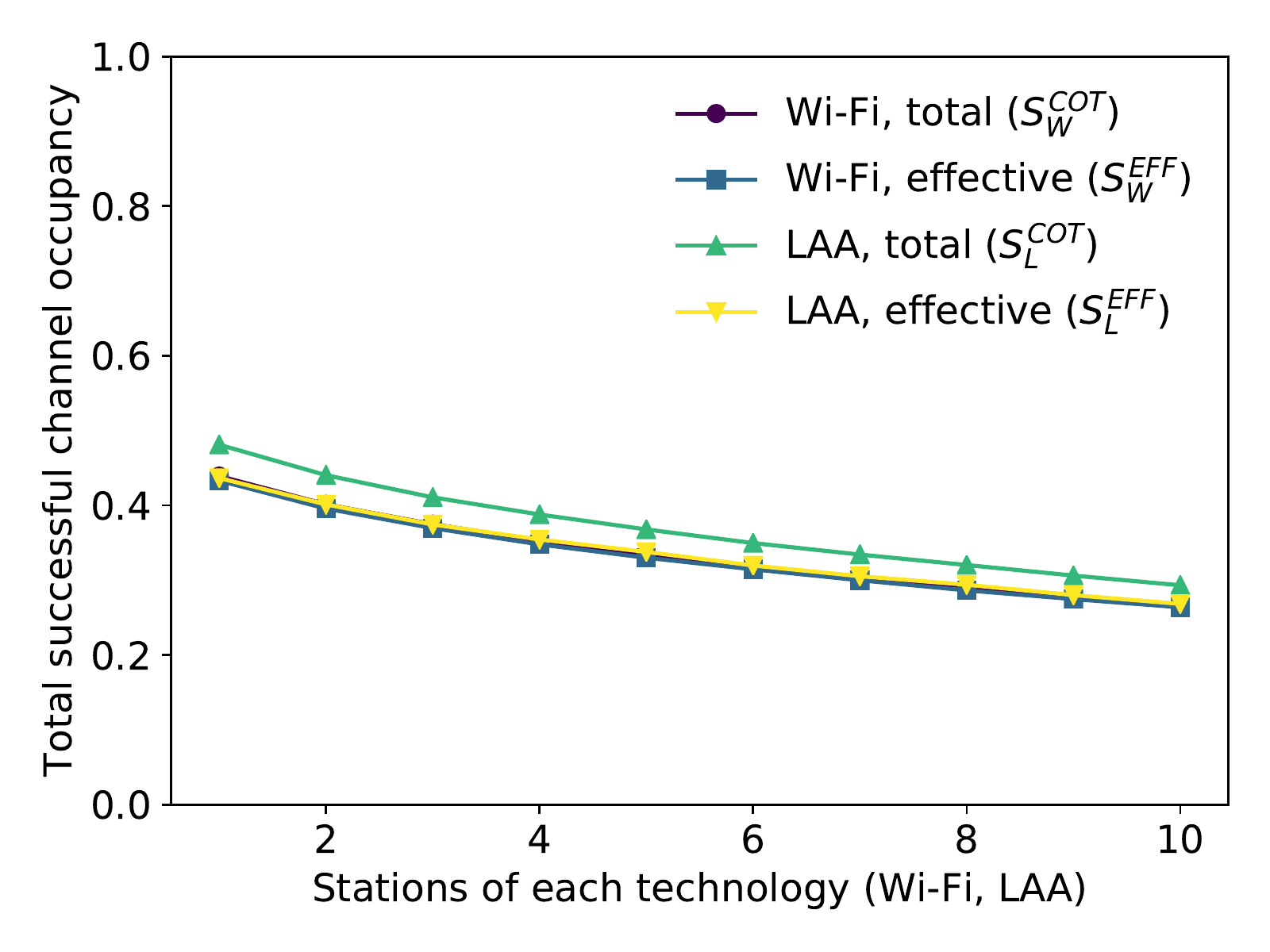}
\label{fig-performance-laa-res-airtime}}
\hfil
\subfloat[Collision probability for RS-based access]{
\includegraphics[width=0.45\textwidth]{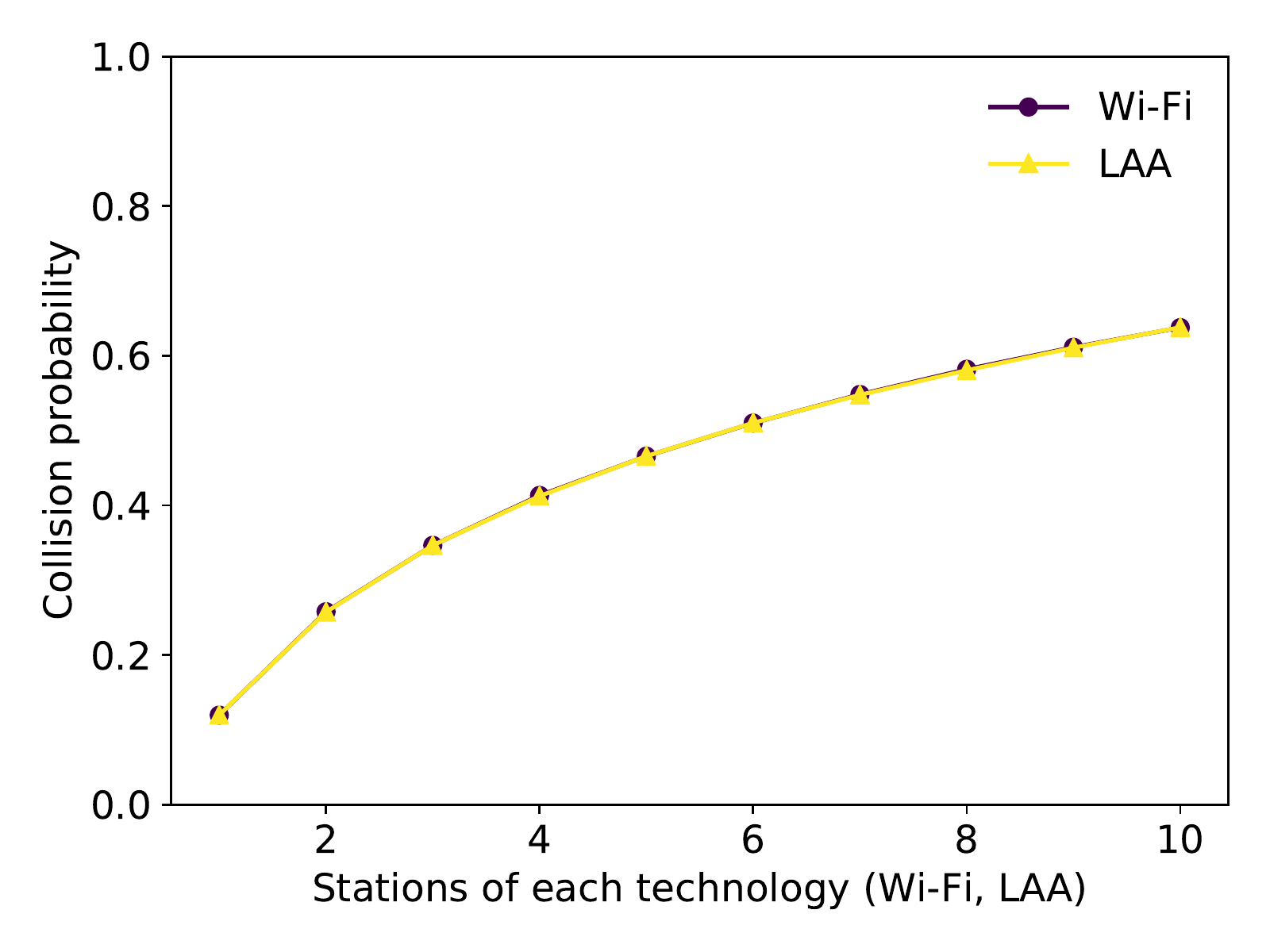}
\label{fig-performance-laa-res-pcol}}
\hfil
\subfloat[Channel occupancy time for gap-based access]{
\includegraphics[width=0.45\textwidth]{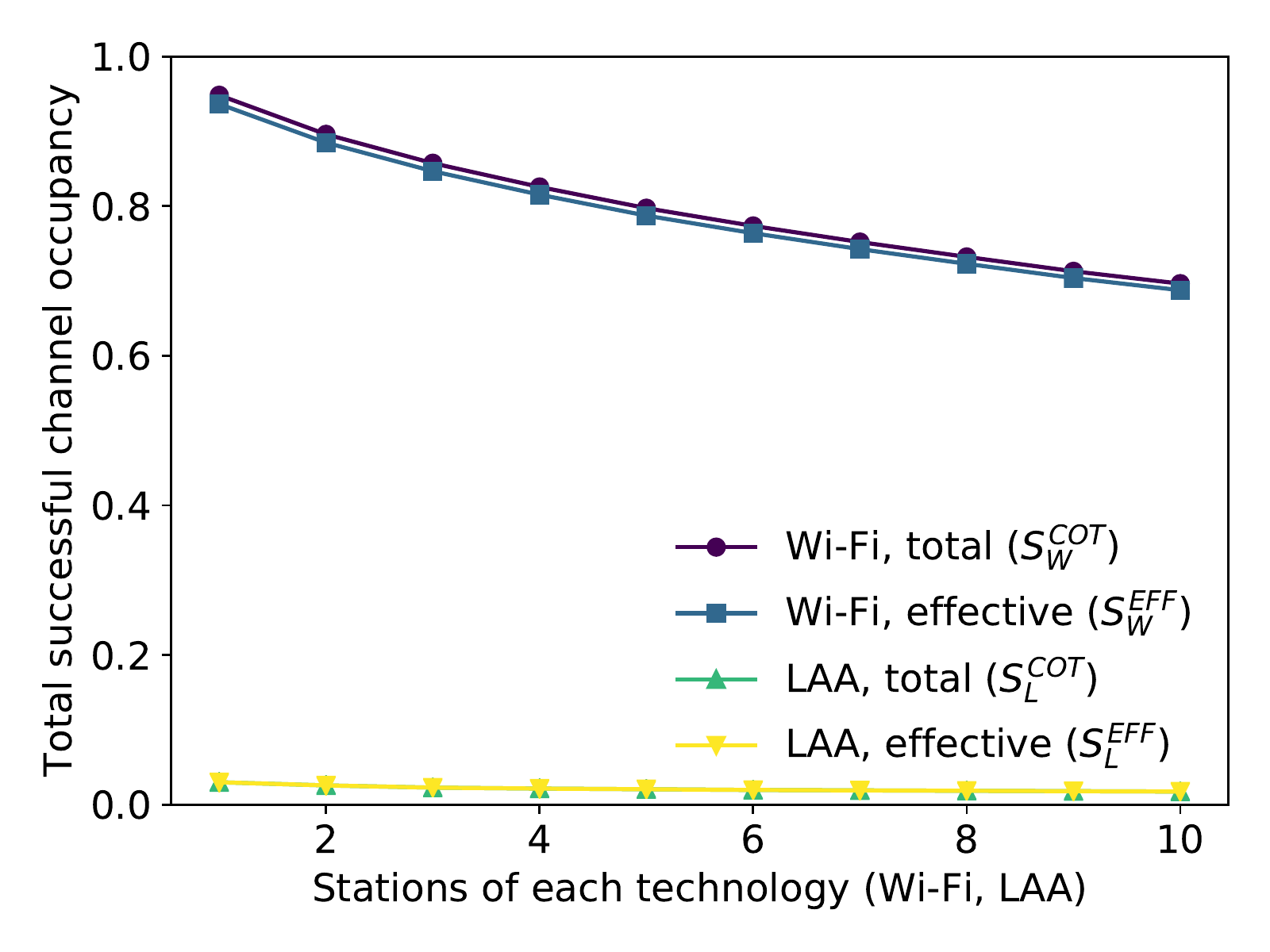}
\label{fig-performance-laa-gap-airtime}}
\hfil
\subfloat[Collision probability for gap-based access]{
\includegraphics[width=0.45\textwidth]{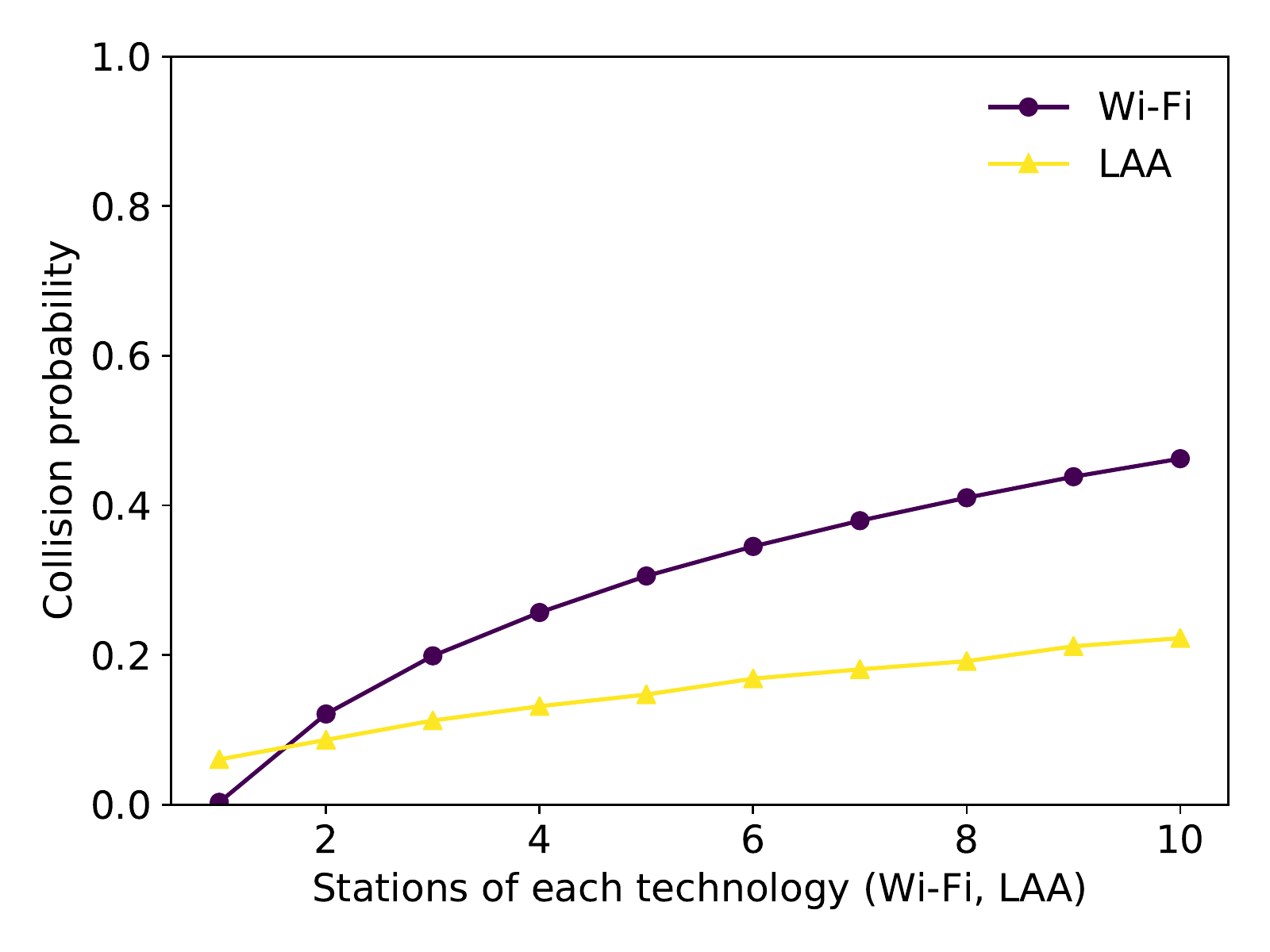}
\label{fig-performance-laa-gap-pcol}}
\hfil
\caption{Performance results for Wi-Fi/LAA: normalized successful total channel occupancy time (left column) and collision probability (right column) when LAA uses the reservation signal (top row) and gap (bottom row) channel access.}
\label{fig-performance-laa}
\end{figure*}

To assess the coexistence performance of Wi-Fi and LAA we measure the successful channel occupancy times $S^{COT}$ and $S^{EFF}$ as well as the collision probability $C$ in two scenarios (LAA with RS-based and LAA with gap-based channel access). 
In both scenarios, all nodes share a single channel, the number of both Wi-Fi and LAA nodes increases from 1 to 10, and  $\Delta$ is set to $1000$~\si{\micro\second} for LAA. 

As shown in Fig.~\ref{fig-performance-laa-res-airtime} and Fig.~\ref{fig-performance-laa-res-pcol}, when LAA nodes use the RS-based approach, both the channel occupancy times and collision probabilities are similar for both technologies \cite{Kutsevol2019Analytical}. 
When the severity of contention increases, the channel occupancy time steadily drops as a result of the higher number of collisions encountered by the transmitting nodes. However, for LAA, part of this time is spent on the reservation signal, and therefore its effective channel occupancy time ($S^\text{EFF}_\text{L}$) is lower than the total one ($S^\text{COT}_\text{L}$).
Therefore, we conclude that the results in this scenario show a slight unfairness towards Wi-Fi.

When LAA nodes implement the gap mechanism, the situation changes significantly, as shown in Fig.~\ref{fig-performance-laa-gap-airtime}. The channel occupancy time drops to near zero for LAA while remaining high for Wi-Fi, leading to severe unfairness. 
Interestingly, the collision probability is also lower for LAA (Fig.~\ref{fig-performance-laa-gap-pcol}): LAA nodes can rarely access the channel, but when they do, they rarely collide. 
This scenario shows that in the absence of RSs, a new approach is required to achieve better fairness in channel access and NR-U addresses this with 
the introduction of flexible slot durations.

\begin{figure*}
\centering
\subfloat[Single Wi-Fi and single NR-U node]{
\includegraphics[width=0.45\textwidth]{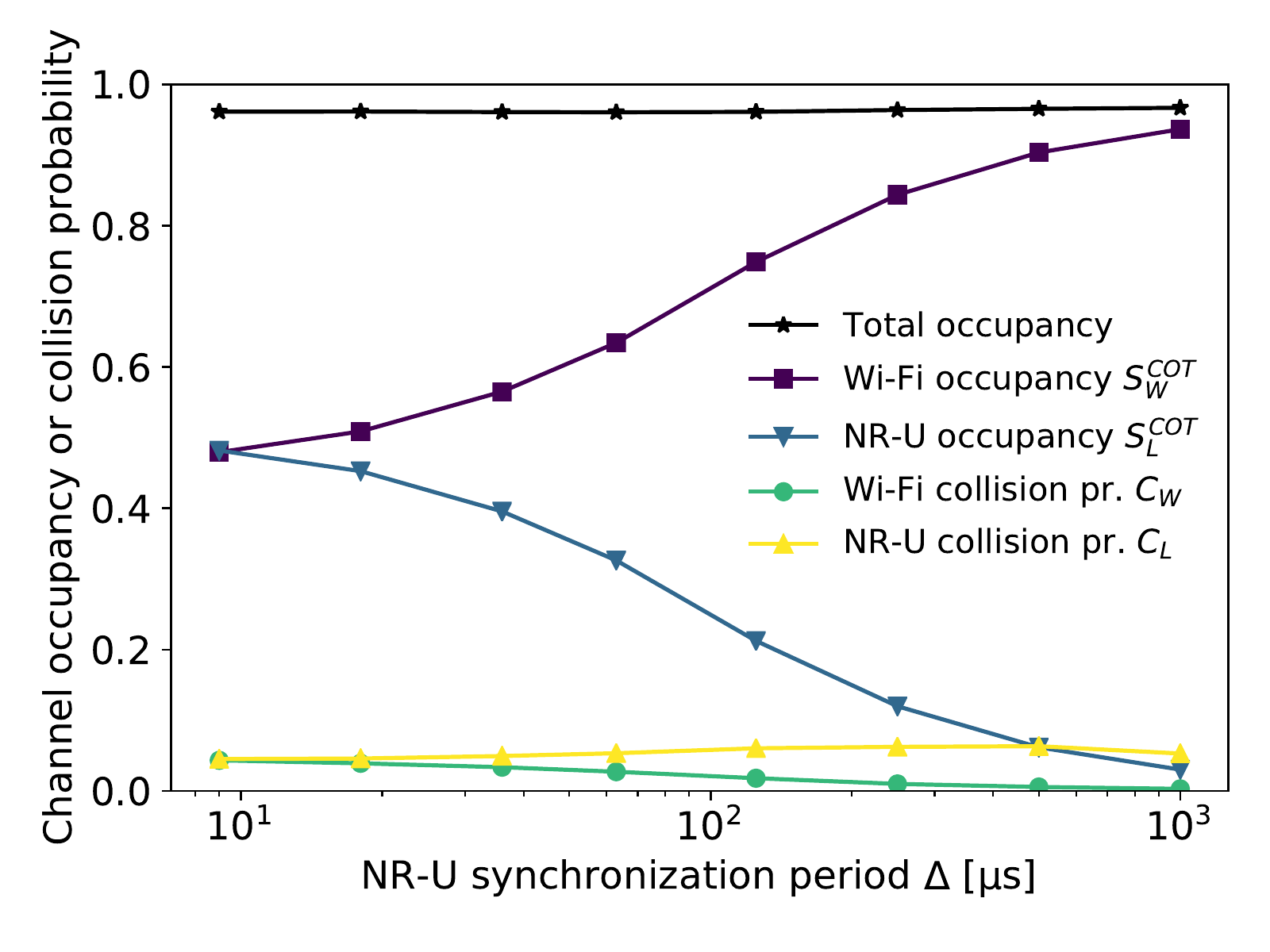}
\label{fig-performance-nr-n=1}}
\hfil
\subfloat[10 Wi-Fi and 10 NR-U nodes]{
\includegraphics[width=0.45\textwidth]{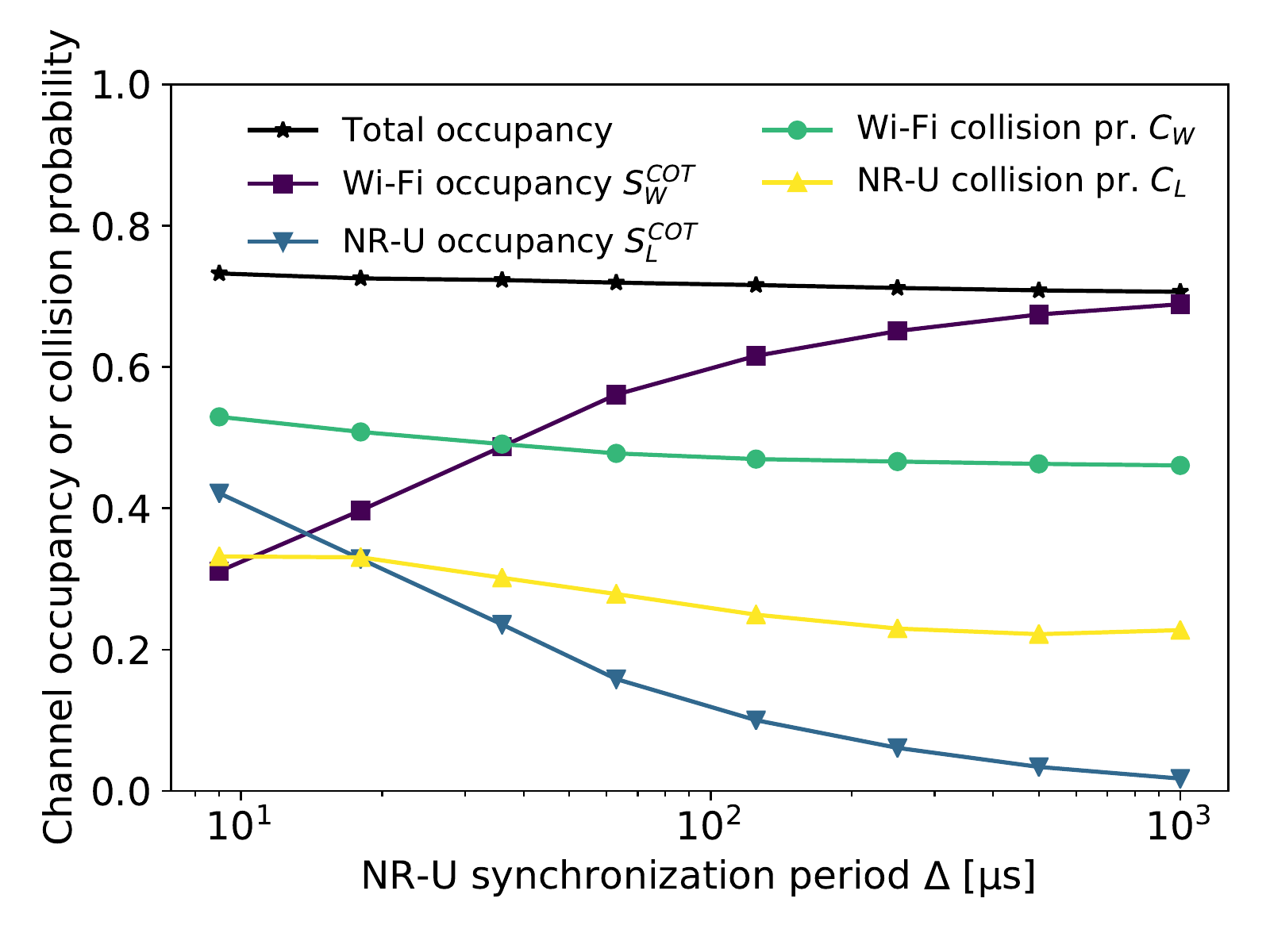}
\label{fig-performance-nr-n=10}}\\
\subfloat[Channel occupancy times]{
\includegraphics[width=0.45\textwidth]{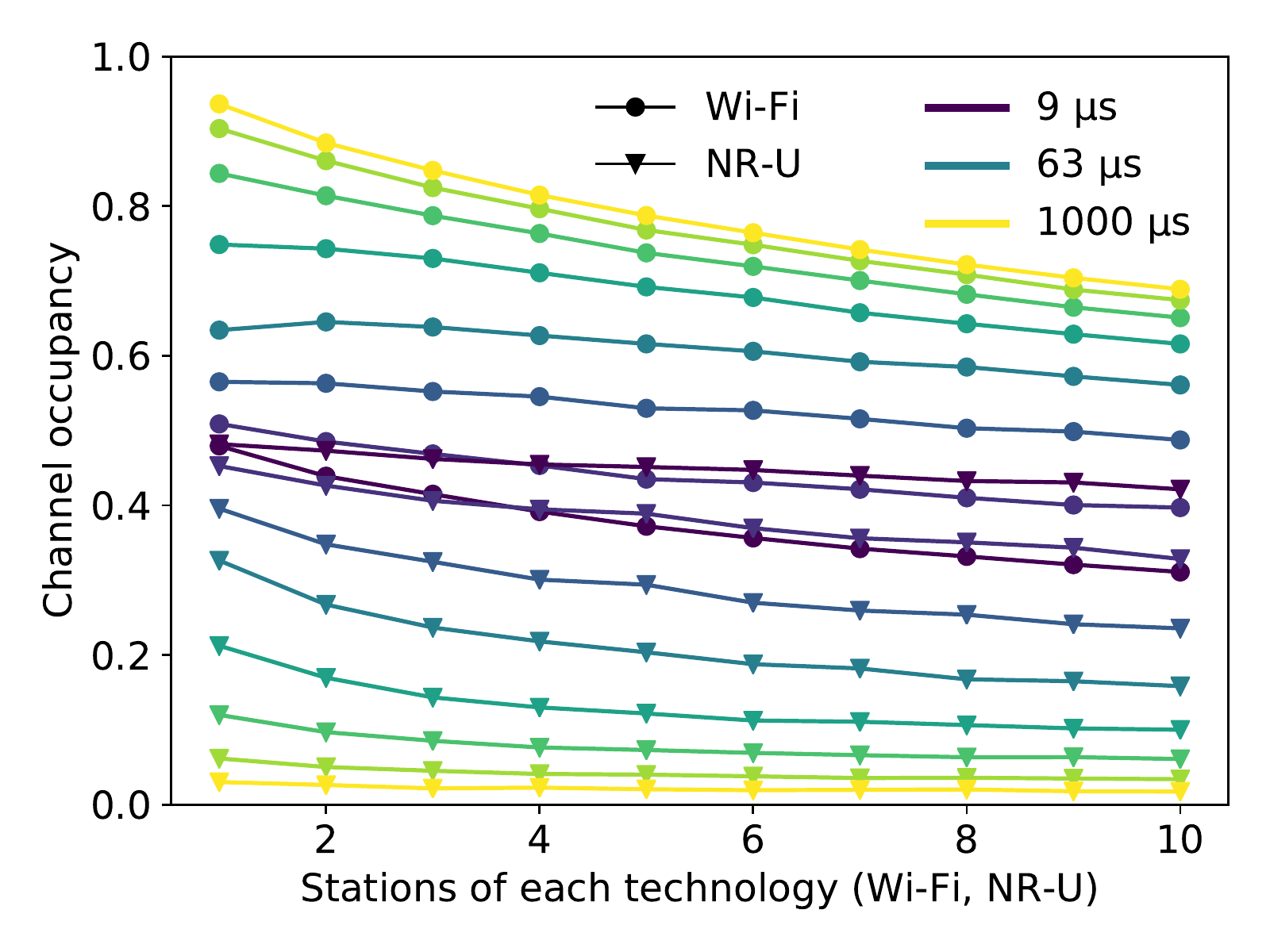}\qquad
\label{fig-performance-nr-comparison}}
\subfloat[Collision probabilities]{
\includegraphics[width=0.45\textwidth]{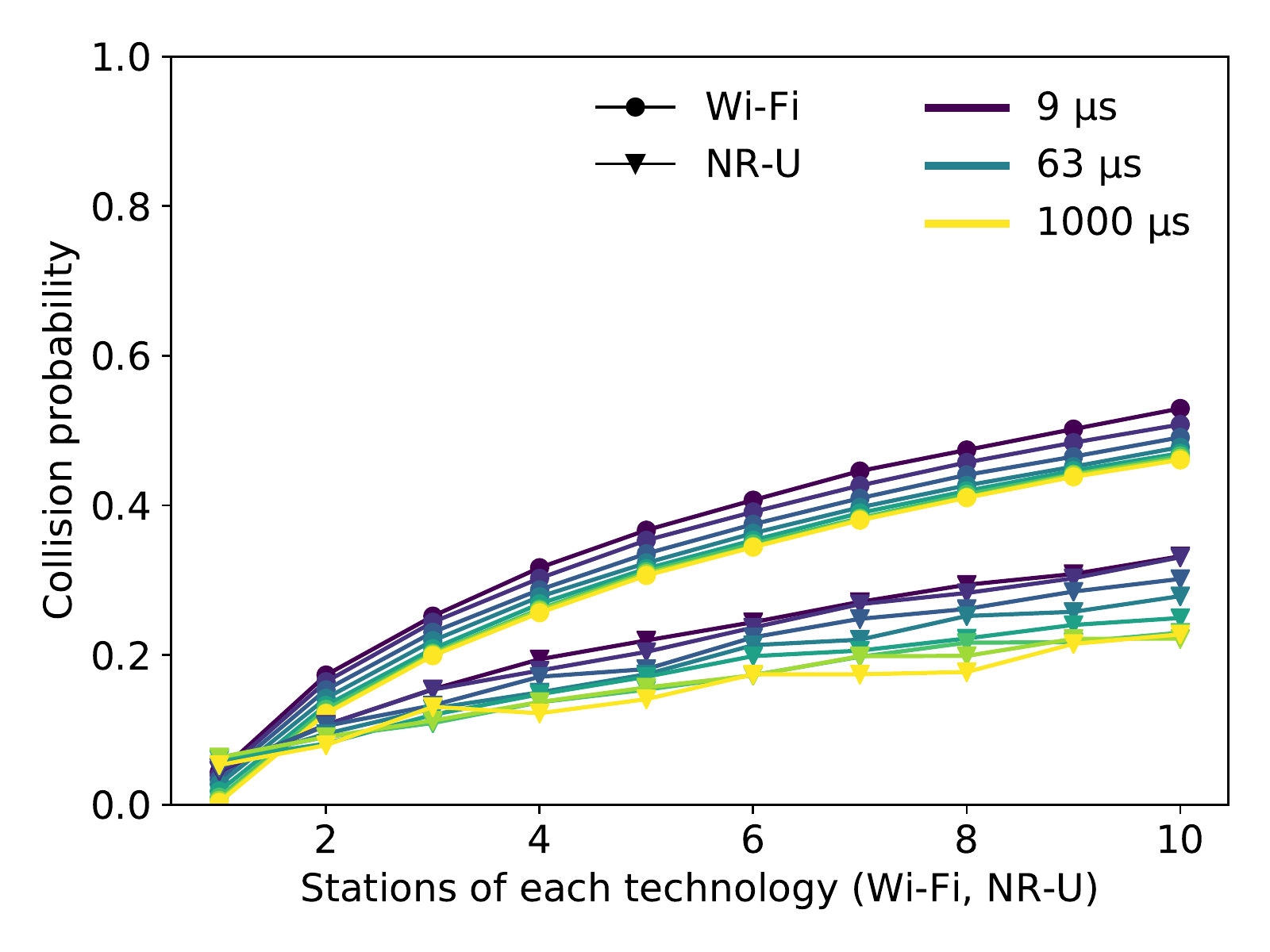}
\label{fig-performance-nr-comparison-pc}}\\
\subfloat[10 Wi-Fi and 10 synchronized NR-U nodes]{
\includegraphics[width=0.45\textwidth]{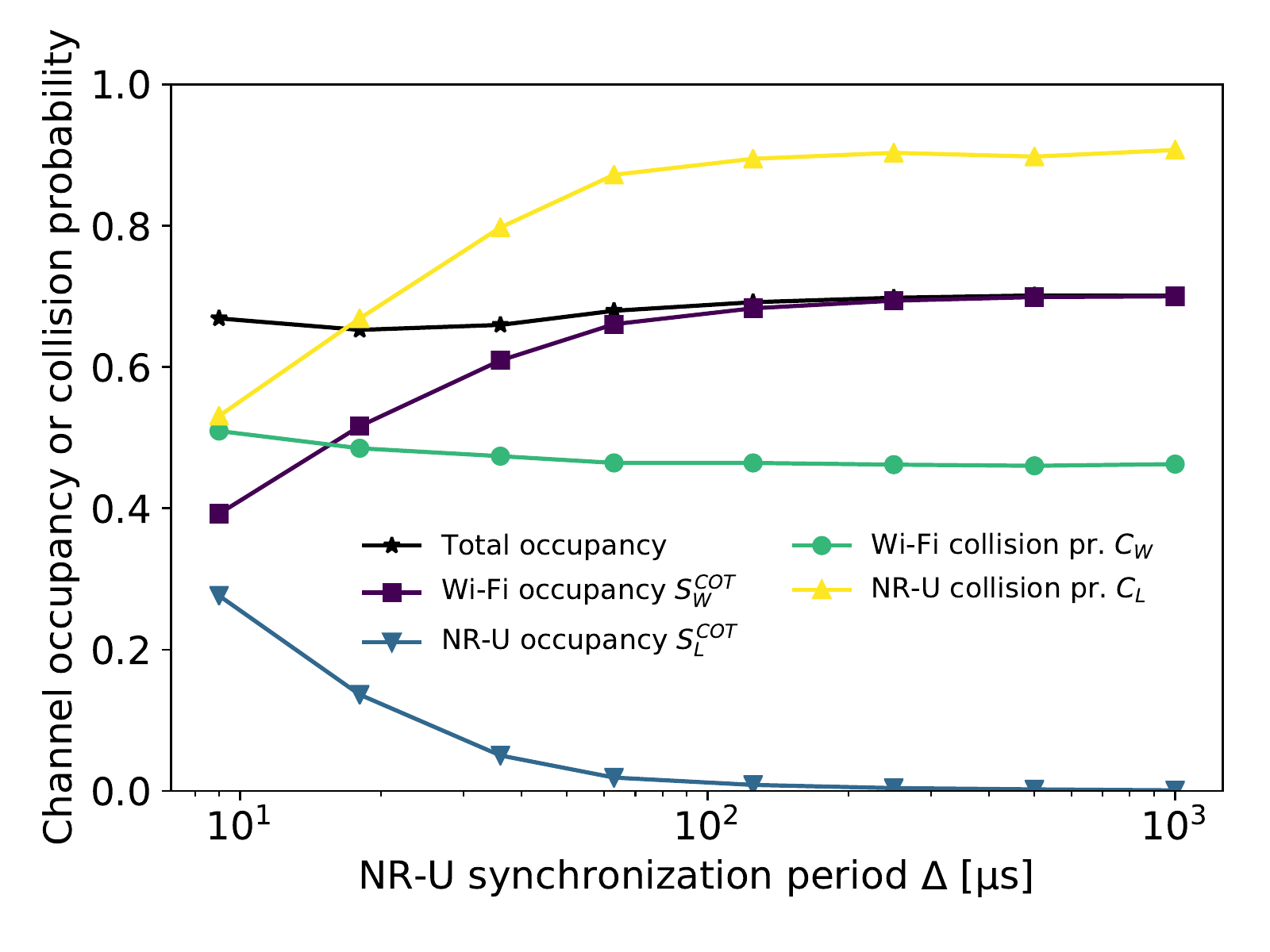}
\label{fig-nr-n=10-noOffset}}
\caption{Performance results for Wi-Fi/NR-U coexistence.}
\label{fig-performance-nr}
\end{figure*}

\subsection{Wi-Fi/NR-U Coexistence}
In the next scenario NR-U and Wi-Fi nodes share a single channel, the number of both Wi-Fi and NR-U nodes is increased from 1 to 10 and $\Delta \in \{9, 18, 32, 63, 125, 250, 500,\allowbreak 1000\}$~\si{\micro\second} for NR-U. The same two performance metrics were measured: successful channel occupancy time  and collision probability. 
Note that for both NR-U and Wi-Fi $S^\text{COT}\approx S^\text{EFF}$, respectively, because there is no RS and the control overhead of Wi-Fi is small compared to the amount of data transmitted.
Therefore, in the figures we plot only $S^{COT}$.

For $\Delta=9$~\si{\micro\second}, when the synchronization slot has the same size as a backoff slot, we expect both Wi-Fi and NR-U to obtain similar results. Indeed, for one Wi-Fi and one NR-U transmitting node, a perfect match is obtained (Fig.~\ref{fig-performance-nr-n=1}), therefore the channel access is completely fair for both technologies. This was obtained without wasting the channel bandwidth for RS transmissions observed for the LAA case. When the number of contenting nodes increases  (Fig.~\ref{fig-performance-nr-comparison}), NR-U slightly outperforms Wi-Fi, e.g., for 20 nodes (Fig.~\ref{fig-performance-nr-n=10}) the normalized channel occupancy time of NR-U is $\simeq$10 percentage points higher. This is because we assume fully desynchronized NR-U nodes, i.e., their synchronization slots are not aligned. 
This means that while Wi-Fi transmissions start at similar times (for the same backoff values), NR-U transmissions start at different times.
When NR-U nodes are perfectly synchronized their collision probability is much higher and the channel occupancy time is considerably lower (Fig.~\ref{fig-nr-n=10-noOffset}). 

For increasing $\Delta$, the normalized channel occupancy time of Wi-Fi increases at the expense of NR-U. For $\Delta=1000$~\si{\micro\second}, the channel occupancy time of NR-U drops almost to zero, regardless of the number of contending nodes (Fig.~\ref{fig-performance-nr-comparison}). This is a result of the low number of won transmission attempts, which also impacts the observed collision probability levels (Fig.~\ref{fig-performance-nr-n=10}). Therefore, in such a case, we conclude that the implemented channel access rules are unfair from the perspective of NR-U and additional mechanisms would be required to provide better fairness to scheduled-access systems.

\subsection{Scheduled and random access nodes: are they good neighbors?}

\begin{figure}[htb]
\centering
\subfloat[]{
\includegraphics[width=\columnwidth]{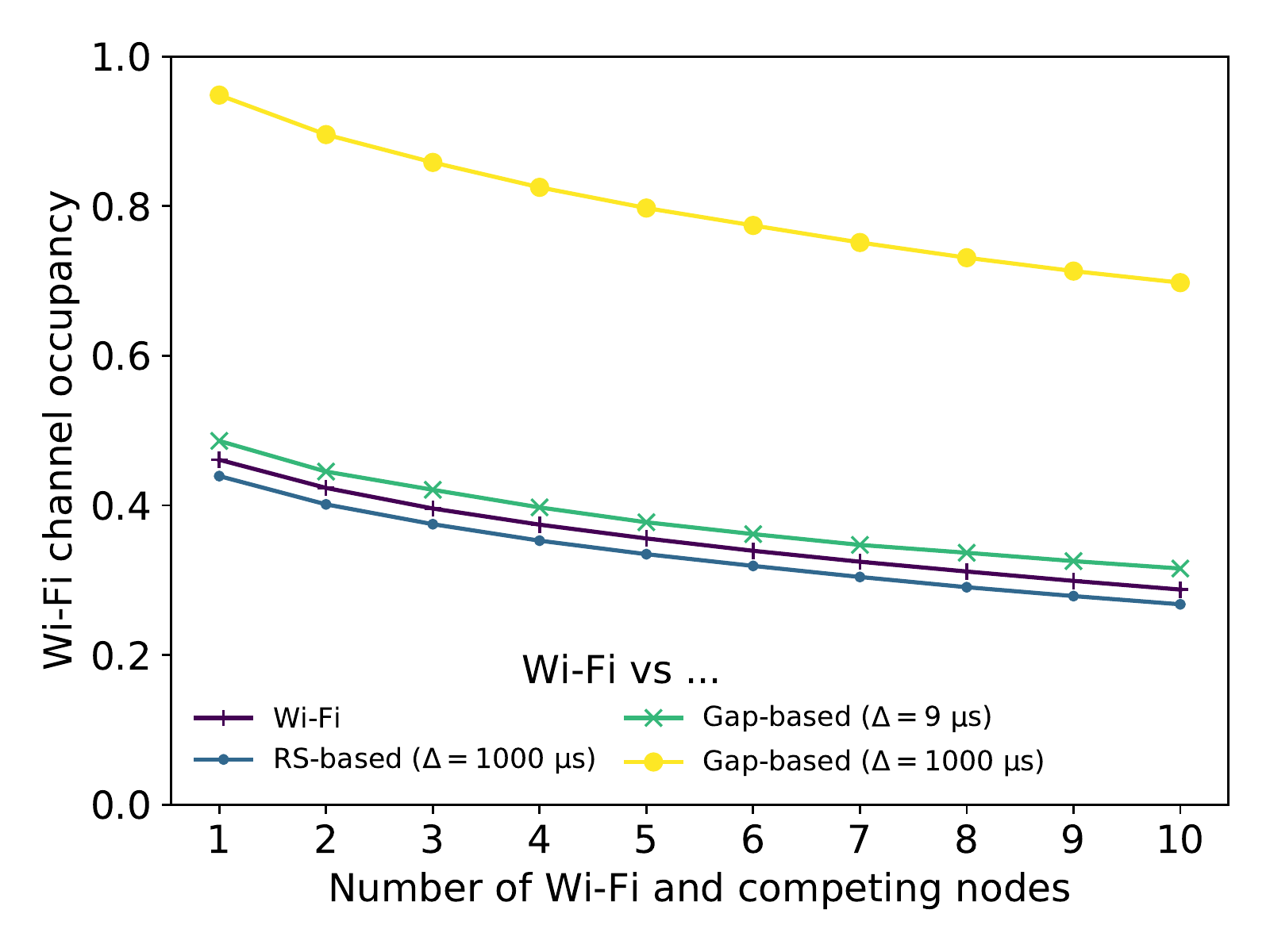}
\label{fig-neighbor-Wi-Fi}}
\hfil
\subfloat[]{
\includegraphics[width=\columnwidth]{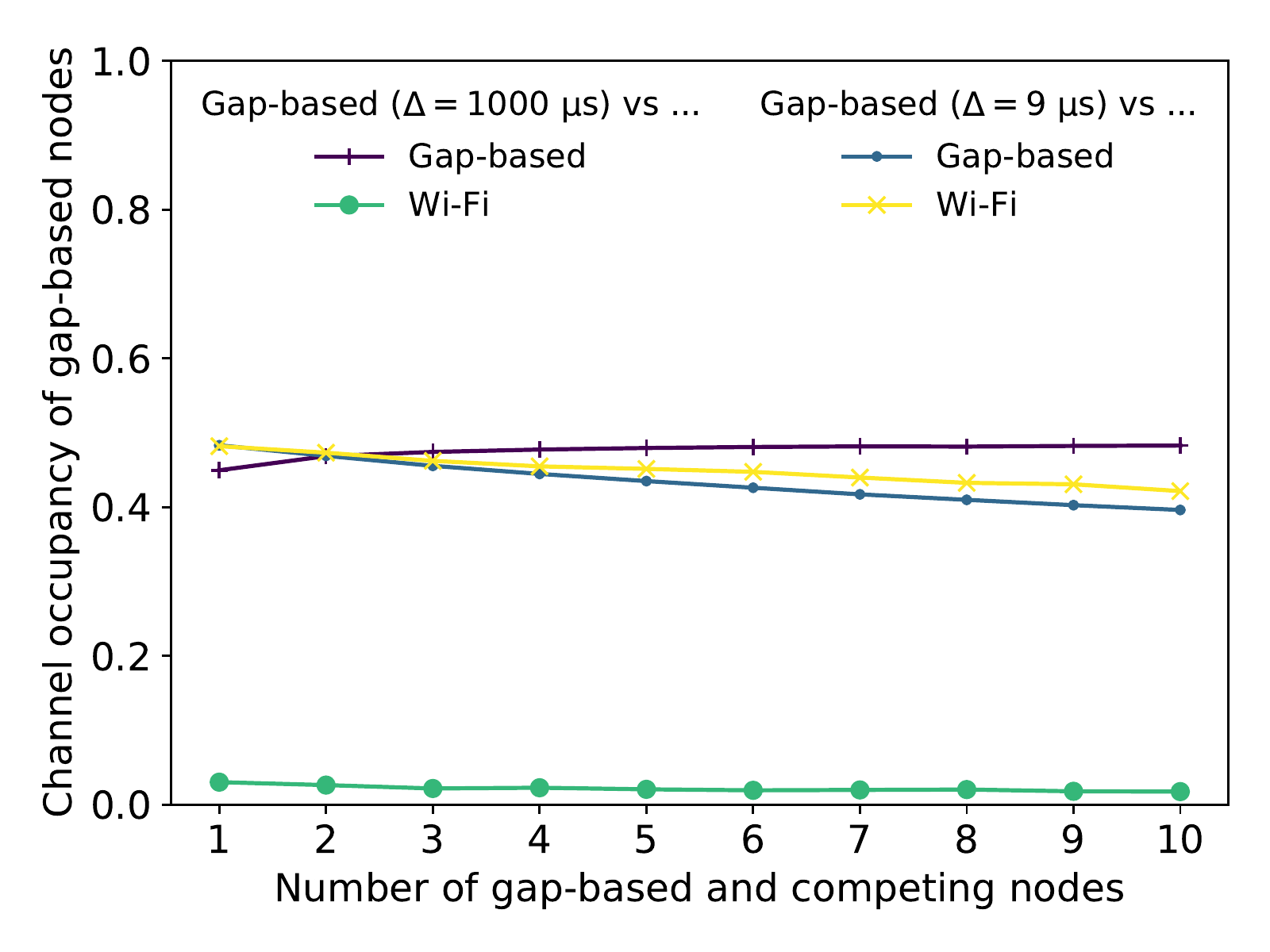}
\label{fig-neighbor-nru}}\\
\caption{Who is a better neighbor for (a) Wi-Fi and (b) gap-based nodes?}
\label{fig-neighbor}
\end{figure}

Next, we analyse the coexistence of nodes implementing random channel access (Wi-Fi) with those using scheduled access (LAA/NR-U) based on either reservation signals or gaps.
For scheduled access we consider extreme synchronization slot values: either $\Delta=1000$~\si{\micro\second} (with RS-based or gap-based) or $\Delta=9$~\si{\micro\second} (only gap-based since RS does not bring gains for such a short slot).
The former configuration setting is available for both LAA and NR-U, while the latter -- only for NR-U.
Furthermore, we again assume maximum channel occupancy values $o_{\max}$ (i.e., 5.4~ms for Wi-Fi and 6~ms for LAA/NR-U) and an evenly increasing number of nodes for each technology (from one to ten). Therefore, in the case with the highest contention there are 20 nodes in total.
The obtained results are presented in Fig.~\ref{fig-neighbor} as the channel access occupancy for, respectively, Wi-Fi and gap-based nodes as a function of the increasing number of competing nodes. 
The baselines in each case are the Wi-Fi curve (Fig.~\ref{fig-neighbor-Wi-Fi}) and the gap-based curves (in Fig.~\ref{fig-neighbor-nru}).

Nodes implementing gap-based access and a synchronization slot of $\Delta=1000$~\si{\micro\second} seem to be ideal neighbors for Wi-Fi (Fig.~\ref{fig-neighbor-Wi-Fi}). Under such a large synchronization slot, the gap mechanism  hinders their opportunity to transmit leaving the channel available to Wi-Fi. However, in this case the channel access is considerably unfair towards gap-based nodes.

Nodes implementing the RS mechanism are slightly worse neighbors for Wi-Fi , because
their transmission duration can be slightly longer (6~ms vs 5.4~ms) according to specifications Table \ref{tab:params}, owing to the out-of-band acknowledgment signalling (Fig.~\ref{fig-neighbor-Wi-Fi}). 
Interestingly, gap-based nodes nodes with 9~\si{\micro\second} synchronization slots are better neighbors to Wi-Fi than if they were replaced by other Wi-Fi nodes, even though $o_{\max}$ for NR-U is slightly longer. This is a result of the lower probability of collision between such nodes.

Nodes implementing  gap-based channel access are best neighbors for each other when using 1000~\si{\micro\second} synchronization slots (Fig.~\ref{fig-neighbor-nru}). However, under a 9~\si{\micro\second} synchronization slot, Wi-Fi nodes are better neighbors to such NR-U nodes, because they have a shorter $o_{\max}$. The advantage of the coexistence with Wi-Fi is obviously even higher when the contending NR-U nodes are synchronized with each other (as shown in Fig.~\ref{fig-nr-n=10-noOffset}).

To summarize, LAA/NR-U (using either RS-based or gap-based channel access) and Wi-Fi nodes are only partially good neighbors to each other. The introduction of flexible slot duration improves the fairness in channel access for the coexisting scheduled and distributed technologies, though only in selected settings. For large synchronization slot duration, channel access can be considered unfair from the perspective of LAA/NR-U nodes. Therefore, novel mechanisms will be required to improve this behaviour in future deployments, such as the one proposed by Qualcomm in \cite{Qualcomm2019}, which requires modifying MCOT for synchronous technologies such as LAA/NR-U.

\section{Conclusions}
In this paper we have analyzed the coexistence of technologies using random (Wi-Fi) and scheduled (LAA/NR-U) channel access in unlicensed bands. We considered two alternative channel access mechanisms for synchronous systems (gap- and reservation signal-based). Additionally, we analyzed the impact of NR-U's scheduling flexibility. The main findings of our work are as follows:
\begin{enumerate}
    \item In terms of modelling, heterogeneous channel access mechanisms cannot rely on the persistent approximation proposed by Bianchi \cite{Bianchi2000}, according to which each node is represented by a channel access probability uniform over time. However, it is still possible to model the system memory as a Markov process, by characterizing the evolution of both the backoff counters and the additional synchronization times (if any) required by the nodes for channel contention. 
    \item When scheduled systems use gap-based channel access:
    \begin{itemize}
        \item they avoid the problems caused by using RSs,
        \item they can achieve airtime fairness with Wi-Fi only when their scheduling is configured to the shortest synchronization slots,
        \item they should not be synchronized with each other, otherwise, under high contention they starve each other,
        \item they are good neighbours to Wi-Fi because they do not waste bandwidth and do not increase Wi-Fi's collision probability (in comparison to an all-Wi-Fi case).
    \end{itemize}
\end{enumerate}
While the performed analysis was done for the 5~GHz band, in our opinion the results can be generalized to all sub-7~GHz bands. Meanwhile, the operation in millimeter wave bands (60~GHz), is specific due to the different propagation characteristics and the use of beamforming. We refer the reader to \cite{Patriciello2020} for an in-depth study of this case.

As future work we envision the analysis of uplink transmissions, which are more complex: the behavior of Wi-Fi for single-user transmissions remains similar, while scheduled systems (LAA/NR-U and Wi-Fi with OFDMA) can show their potential by better avoiding intra-network collisions. Additionally, we plan to propose novel mechanisms (such as usage of adaptive contention windows and inter-frame spaces) to provide fair coexistence of scheduled and distributed systems in the unlicensed bands. Other possible areas of study are QoS (the interaction between different traffic priorities), new 802.11ax features, such as OFDMA-based channel access, which bring Wi-Fi closer in principles of operation to scheduled systems, the impact of LAA/NR-U's multi-channel access \cite{sathya2020standardization},
as well as a comparison between gap-based access and alternative RS-based channel access methods \cite{Loginov2021}.

\section*{Acknowledgment}

The work of K. Kosek-Szott and S. Szott was carried out as part of a project financed by the Polish National Science Centre (DEC-2018/30/M/ST7/00351).

\end{document}